\documentclass[aps,pra,twocolumn,showpacs,showkeys,amsmath,superscriptaddress,floatfix,longbibliography,nofootinbib]{revtex4-1}
\usepackage{csquotes, relsize}
\usepackage{amsmath}
\usepackage{amssymb}
\usepackage{makeidx}
\usepackage{amsfonts}
\usepackage[utf8]{inputenc}
\usepackage[T1]{fontenc}
\usepackage[usenames,dvipsnames]{pstricks}
\usepackage{subfigure}
\usepackage{epsfig}
\usepackage{pst-grad} 
\usepackage{pst-plot} 
\usepackage[colorlinks,hyperindex]{hyperref}
\hypersetup
{
	colorlinks,%
	citecolor=black,%
	linkcolor=black,%
	urlcolor=black,%
}

\renewcommand{\vec}[1]{\mathbf{#1}}

\begin{document}
	
	\title{Pure Goldstone mode in the quench dynamics\\of a confined ultracold Fermi gas in the BCS-BEC crossover regime}
	
	\author{P.~Kettmann}
	\affiliation{Institut f\"ur Festk\"orpertheorie, Westf\"alische
		Wilhelms-Universit\"at M\"unster, 48149 M\"unster,
		Germany}
		
	\author{S.~Hannibal}
	\affiliation{Institut f\"ur Festk\"orpertheorie, Westf\"alische
		Wilhelms-Universit\"at M\"unster, 48149 M\"unster,
		Germany}
	
	\author{M.~D.~Croitoru}
	\affiliation{Theoretische Physik III, Universit\"at Bayreuth, 95440
		Bayreuth,
		Germany}
		
	\author{V.~M.~Axt}
	\affiliation{Theoretische Physik III, Universit\"at Bayreuth, 95440
		Bayreuth,
		Germany}
	
	\author{T.~Kuhn}
	\affiliation{Institut f\"ur Festk\"orpertheorie, Westf\"alische
		Wilhelms-Universit\"at M\"unster, 48149 M\"unster,
		Germany}
	
	\date{\today}
	
	\begin{abstract}

We present a numerical study of the dynamic response of a confined superfluid Fermi gas to a rapid change of the scattering length (i.e., an interaction quench). Based on a fully microscopic time-dependent density-matrix approach within the full Bogoliubov-de Gennes formalism that includes a 3D harmonic confinement we simulate and identify the emergence of a Goldstone mode of the BCS gap in a cigar-shaped $^6$Li gas.
By analyzing this Goldstone mode over a wide range of parameters, we show that its excitation spectrum is gapless and that its main frequency is not fixed by the trapping potential but that it is determined by the details of the quench. Thus, we report the emergence of a pure Goldstone mode of the BCS gap that --in contrast to situations in many previous studies-- maintains its gapless excitation spectrum predicted by the Goldstone theorem. Furthermore, we observe that the size-dependent superfluid resonances resulting from the atypical BCS-BEC crossover have a direct impact on this Goldstone mode. Finally, we find that the interaction quench-induced Goldstone mode leads to a low-frequency in-phase oscillation of the single-particle occupations with complete inversion of the lowest-lying single-particle states which could provide a convenient experimental access to the pure gapless Goldstone mode.
	\end{abstract}
	


	\maketitle
	
	\section{Introduction}

\maketitle
Due to their unique controllability ultracold Fermi gases provide an ideal system to test concepts of many-particle physics as well as particle theory.  Adjustable interparticle interactions provide the possibility to explore both the regime of weak attractive interactions where a superfluid Bardeen Cooper Schrieffer (BCS) phase emerges as well as the regime of weak repulsive interactions which lead to the formation of a Bose-Einstein condensate (BEC). Both regimes are connected by a smooth BCS-BEC crossover with strong interparticle interactions including a point of unitarity where the coupling strength diverges \cite{Giorgini2008Theory, Bloch2008Many}. Furthermore, the emergence of a BCS phase is associated with a spontaneously broken $U(1)$ symmetry which makes ultracold Fermi gases a convenient candidate to study the fundamental concept of spontaneous symmetry breaking (SSB) \cite{weinberg1996quantum}.

\begin{figure}[t]
	\centering
	\includegraphics[width= 0.64 \columnwidth]{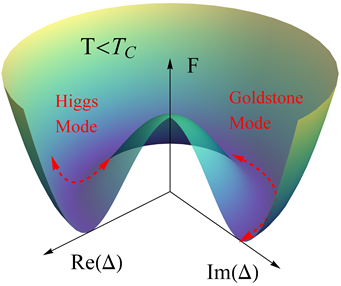}
	\caption{(color online) Ginzburg-Landau free energy of a BCS phase of a homogeneous Fermi gas for $T<T_C$ (schematic). The rotational $U(1)$ symmetry is broken, when the gap $\Delta$ of the system gains a certain phase inside the rim of the potential. The resulting excitation modes are the Higgs-amplitude mode and the Goldstone mode, i.e., a phase oscillation of the gap.}
	\label{fig:Mexican}
\end{figure}

Spontaneously broken gauge symmetries and the resulting two types of fundamental collective excitations --gapped amplitude/Higgs modes and gapless phase/Goldstone modes (see Fig. \ref{fig:Mexican})-- are of fundamental interest for several fields of physics like condensed matter and particle physics. Probably the most prominent application of the concept of SSB is the Higgs mechanism in particle physics \cite{Higgs1964Broken}. In condensed matter physics SSB occurs in several systems, for example in ferromagnets (see, e.g., \cite{burgess2000goldstone}), superfluid $^3$He \cite{Paulson1973Propagation,Lawson1973Attenuation} and BCS superconductors \cite{anderson1958random, weinberg1996quantum}. In these cases, the fundamental excitations --known as magnons (i.e., spin waves), second sound (i.e., heat waves) and plasmons-- correspond to the Goldstone modes resulting from SSB.

In the field of ultracold Fermi gases the Higgs and the Goldstone mode have received great attention over the past years. The Higgs mode, i.e., the amplitude oscillation of the BCS gap, is difficult to address in experiment since it does not couple directly to external probes \cite{pekker2015amplitude}. This is why measurements of the Higgs mode have only recently been achieved for lattice superfluids \cite{bissbort2011detecting,Endres2012The/Higgs/amplitude} (in the case of BCS superconductors an experimental access has been found via THz spectroscopy \cite{Matsunaga2013Higgs,Matsunaga2014Light}), while several theoretical studies on the Higgs mode have been reported \cite{Barankov2004Collective,Barankov2006Synchronization,Yuzbashyan2006Relaxation,Dzero2007Spectroscopic,Scott2012Rapid,Bruun2014Long,hannibal2015quench}. In contrast, the Goldstone mode, i.e., the oscillation of the complex phase of the BCS gap, has been intensively studied both theoretically and experimentally (see, e.g., \cite{Kinast2004Evidence, Kinast2004Breakdown, bartenstein2004Collective, altmeyer2007dynamics,  altmeyer2007precision, riedl2008collective, baranov2000low, bruun2001low, bruun2002low, hu2004collective, heiselberg2004collective, Stringari2004Collective, grasso2005temperature, korolyuk2011density}).

However, most previous studies were based on the fact that the Goldstone mode couples to the real-space dynamics of the Fermi gas. Therefore, it can be excited by inducing a collective oscillation of the trapped cloud, which can be achieved, e.g., by various schemes of confinement change or by optical excitation. The dynamics in the Goldstone mode is then directly reflected in the collective oscillation of the cloud and can be observed via the latter. However, a coupling to the real-space oscillation of the cloud implies a coupling to the trapping potential. I.e., in the previous studies the frequency of the Goldstone mode was fixed by the frequencies of the trap which stands in direct contrast to the Goldstone theorem predicting a gapless excitation spectrum of the phase mode. Thus, the observation of the gapless phase mode resulting from the Goldstone theorem was so far obstructed by the coupling to the trapping potential via the collective oscillation of the cloud. 

In contrast, we report the emergence of a pure Goldstone mode in the dynamics of a trapped ultracold Fermi gas by showing that its original gapless excitation spectrum can be recovered by exciting the cloud via an interaction quench. To this end, we observe that the frequency of the interaction quench-induced phase mode can be tuned over a wide range from zero to finite values by adjusting the details of the quench. We explain this by the circumstance that the excited Goldstone mode is homogeneous, i.e., that it does not introduce any phase gradients, and that it therefore does not couple to the trapping potential.

To do so, we study the dynamics of the BCS gap of a confined ultracold $^6$Li gas at $T=0$ on the BCS side of the BCS-BEC crossover [i.e., the Fermi wave vector $k_F$ times the scattering length $a$ is given by $0>1/(k_Fa)>-1$] as well as in the BCS regime [i.e., $-1>1/(k_Fa)$]. We calculate the dynamics of the Bogoliubov quasiparticles in the framework of the Bogoliubov-de Gennes (BdG) formalism and by that the dynamics of the BCS gap. The investigated system is in the low-density regime, i.e., we use a short-range s-wave interaction between particles with opposite spin. The confinement is modeled by a cigar-shaped 3D harmonic potential, which in good approximation describes the standard laser confinement used in experiment \cite{Bloch2008Many}. The system is excited by an instantaneous interaction quench, i.e., a rapid change of the interparticle interaction strength. This can experimentally be achieved, e.g., by an optical control of a Feshbach resonance \cite{clark2015quantum}.

Our work is structured as follows. In section \ref{sec:formalism} we present the formalism we used within the context of this work, i.e., a full BdG approach as well as its simplification via the Anderson approximation. We will derive the equations of motion for the quasiparticle expectation values which allow for a calculation of the phase dynamics of the BCS gap. In section \ref{sec:results} we present the results obtained by the full BdG approach and show that they are well reproduced by Anderson's approximate solution. Based on the latter we will analyze the effect of the excitation parameters and of the trapping frequencies on the interaction-quench-induced Goldstone mode as well as the impact of this Goldstone mode on an experimentally accessible quantity, i.e., the single-particle excitations of the cloud. Finally, we conclude and summarize our findings in section \ref{sec:conclusions}.

	\section{Theoretical approach}\label{sec:formalism}
	
	To calculate the ground state as well as the dynamical properties of a BCS condensate in a trap we start from the Bogoliubov-de Gennes (BdG) Hamiltonian \cite{DeGennes1989Superconductivity,Datta1999Can}:
	\begin{align}\label{eq:BDG}
	H_{BdG} &= \int \, \Psi_{\uparrow}^{\dagger}(\vec{r}) H_0 \Psi_{\uparrow}^{}(\vec{r})
	\,  \mathrm{d}^3r + \int \,   \Psi^{\dagger}_{\downarrow}(\vec{r}) H_0 \Psi^{ }_{\downarrow}(\vec{r})
	\,\mathrm{d}^3 r \notag \\
	&+  \int   \, \left[ \Delta(\vec{r}) \Psi_{\uparrow}^{\dagger}(\vec{r})
	\Psi_{\downarrow}^{\dagger}(\vec{r}) + \Delta^*(\vec{r}) \Psi_{\downarrow}^{ }(\vec{r})
	\Psi _{\uparrow}^{ }(\vec{r}) \right] \, \mathrm{d}^3 r.
	\end{align}
Here the field operators $\Psi^{(\dagger)}_{\sigma}(\vec{r})$ describe the annihilation (creation) of Fermions --in this case atoms of $^6$Li-- with spin $\sigma$ at the position $\vec{r}$ and the BCS gap is given by
\begin{equation} 
\Delta(\vec{r}) = - g \left< \Psi_{\downarrow}(\vec{r}) \Psi_{\uparrow}(\vec{r}) \right>.
\end{equation}
The interaction strength $g = - \frac{4 \pi \hbar^2 a}{m}$ is determined by the s-wave scattering length $a$ and the mass of the particles $m$. The 3D harmonic trapping potential $V(\vec{r})$ is included in the one-particle Hamiltonian $H_0=\frac{p^2}{2m}+V(\vec{r})- \mu$ with the chemical potential $\mu$, where we set the trapping frequencies to $f_x=f_y=:f_\perp \gg f_\parallel := f_z$, i.e., a cigar-shaped trap.\\
In the following, we will show how to diagonalize Eq. \eqref{eq:BDG} and thus obtain its eigenstates and energies which describe the single-particle excitations of the BCS condensate.\\

\subsection{The BdG ground state}\label{ssec:groundstate}

To calculate the ground-state properties of the BCS condensate it is instructive to write the eigenvalue equation corresponding to Eq. \eqref{eq:BDG} as the BdG equation \cite{DeGennes1989Superconductivity}
	\begin{equation}\label{eq:Bogoliubov-Gl}
	\begin{pmatrix} H_0 & \Delta(\vec{r}) \\ \Delta^*(\vec{r}) & -H_0^* \end{pmatrix}
	\left(\begin{array}{c} u_M(\vec{r}) \\ v_M(\vec{r}) \end{array} \right)
	= E_M \left( \begin{array}{c} u_M(\vec{r}) \\ v_M(\vec{r}) \end{array} \right).
	\end{equation}
This equation has the form of a one-particle Schrödinger equation\footnote{To be precise, the many-body nature of the BCS pairing is included in Eq. \eqref{eq:Bogoliubov-Gl} via the self-consistent calculation of the BCS gap $\Delta(\vec{r})$ (see below).}, which implies that $H_{\text{BdG}}$ describes non-interacting quasiparticles, i.e., the single-particle excitations of the BCS condensate. Therefore, Eq. \eqref{eq:Bogoliubov-Gl} can be diagonalized which yields the corresponding single-particle wave functions $\left[ u_M(\vec{r}), v_M(\vec{r}) \right]$ and energies $E_M$. However, before we do so, we want to state that the single-particle states divide into two branches (one of positive energies $E_M = E_{m\alpha} > 0$ and one of negative energies $E_{M} = E_{m\beta} = -E_{m\alpha}$) that can be expressed by one another \cite{Datta1999Can}. Accordingly, we simplify our formalism by writing the corresponding expressions --whenever possible-- solely in terms of the positive-energy states and by dropping the index $\alpha$. I.e., in the following we set $\left[u_m(\vec{r}), v_m(\vec{r})\right]:=\left[u_{m\alpha}(\vec{r}), v_{m\alpha}(\vec{r})\right]$ and $E_m:=E_{m\alpha}$. Furthermore, we transform into the excitation picture, i.e., we flip the $\beta$-branch $\left[(m\beta) \rightarrow (mb)\right]$ and leave the $\alpha$-branch unchanged $\left[(m\alpha) = (ma)\right]$. This yields one twofold degenerate branch with $E_m=E_{ma}=E_{mb}$. The corresponding creation operators read
\begin{equation}\label{eq:B-Transformation1}
	\gamma_{ma}^{\dagger} = \int \, \left[  u_{m}(\vec{r}) \Psi_{\uparrow}^{\dagger}(\vec{r}) + v_{m}(\vec{r}) \Psi_{\downarrow}^{}(\vec{r}) \right] \, \mathrm{d}^3 r
\end{equation}
\begin{equation}\label{eq:B-Transformation2}
	\gamma_{mb}^{\dagger} =  \int \, \left[ u_{m}(\vec{r}) \Psi_{\downarrow}^{\dagger}(\vec{r}) -v_{m}(\vec{r}) \Psi_{\uparrow}^{}(\vec{r})  \right] \, \mathrm{d}^3 r. 	
\end{equation}
We solve the BdG equation by expressing the single-particle states $\left[u_M(\vec{r}), v_M(\vec{r})\right]$ in terms of the bare atomic states, i.e., the eigenstates of the harmonic trap $\phi_i(\vec{r})$,
\begin{eqnarray} 
u_M(\vec{r}) &=\mathlarger{\sum}\limits_{i=1}^N u_M^{(i)} \phi_i(\vec{r}) \label{eq:entw1} \\
v_M(\vec{r}) &=\mathlarger{\sum}\limits_{i=1}^N v_M^{(i)} \phi_i(\vec{r}) \label{eq:entw2}, 
\end{eqnarray}
with $M \in \{ma, mb\}$. Here, we restrict the sum to atomic states from a window of width $\Delta\epsilon \sim 1 \mu$ around the chemical potential (i.e., $0.5\mu < \epsilon < 1.5 \mu$) to reduce the numerical effort\footnote{We have checked that the qualitative features investigated in this work are independent from the size of this window. However, the features shift quantitatively, e.g., the gaps and frequencies shift to larger values when increasing the window size.}.\\
Inserting this in the BdG equation, multipliying by $\phi_m(\vec{r})$ and integrating over $\vec{r}$ yields:
\begin{widetext}
\begin{equation}
\begin{pmatrix}
\epsilon_1 -\mu 		& 0 					& \cdots 	& 0 						& (\Delta)_{11} 	& \cdots 			&  				& (\Delta)_{1N} \\ 
0 						& \epsilon_2-\mu 	& \ddots 	&  \vdots 				& \vdots 			&  \ddots 		&	  				& \vdots \\ 
\vdots 				& \ddots 			& \ddots 	& 0 						& 						&  					&  				&  \\ 
0 						&\cdots  			& 0 			& 	\epsilon_N-\mu 		& (\Delta)_{N1} 	& \cdots 			& 					& (\Delta)_{NN} \\ 	
(\Delta)^*_{11} 	&  	\cdots 		&  			& (\Delta)^*_{1N} 	& -\epsilon_1 +\mu 	& 0 					& \cdots 		& 0 \\ 
\vdots 				& \ddots 			&  			& 	 \vdots 			& 0 					& -\epsilon_2+\mu 	&  	\ddots 	&  \vdots\\ 
 						&  					&  			&	  						&  \vdots 			&  		\ddots 	& \ddots 		& 0 \\ 
(\Delta)^*_{N1} 	& \cdots 			&  			& (\Delta)^*_{NN} 	& 0 					&  	\cdots 		& 0 				&  -\epsilon_N +\mu
\end{pmatrix} 
\begin{pmatrix}
u^{(1)}_M 	\\ 
u^{(2)}_M	\\ 
\vdots		\\ 
u^{(N)}_M \\ 
v^{(1)}_M	\\ 
v^{(2)}_M	\\ 
\vdots 		\\ 
v^{(N)}_M
\end{pmatrix} 
= E_M \begin{pmatrix}
u^{(1)}_M \\ 
u^{(2)}_M \\ 
\vdots 		\\ 
 u^{(N)}_M	\\ 
v^{(1)}_M	\\ 
v^{(2)}_M 	\\ 
\vdots 		\\ 
v^{(N)}_M
\end{pmatrix}, \label{eq:Matrix}
\end{equation}
\end{widetext}
with $(\Delta)_{mn} := \int \mathrm{d}^3 r \phi_m^*(\vec{r}) \Delta(\vec{r}) \phi_n(\vec{r})$.\footnote{To calculate matrix elements of the form $\int \mathrm{d}^3 r \phi_m (\vec{r})\phi_n (\vec{r})\phi_k (\vec{r})\phi_l (\vec{r})$ we used an analytical expression derived in \cite{lord1949some}.} However, the BCS gap can be expressed in terms of the Bogoliubov transformation [Eqs. \eqref{eq:B-Transformation1} and \eqref{eq:B-Transformation2}] which yields
\begin{align}\label{eq:Delta_Bogolon}
\Delta(\vec{r}) = - g  \sum_{m,n}  & v_m^*(\vec{r}) u_n(\vec{r}) \Big<
\gamma_{ma}^{\dagger}\gamma_{na}^{\phantom{\dagger}}\Big> \notag \\
+ & u_m(\vec{r}) u_n(\vec{r}) \left< \gamma_{mb}^{\phantom{\dagger}}
\gamma_{na}^{\phantom{\dagger}} \right> \notag \\
- & v_m^*(\vec{r}) v_n^*(\vec{r}) \left< \gamma_{ma}^{\dagger}\gamma_{nb}^{\dagger}
\right> \notag \\
+ & u_m(\vec{r}) v_n^*(\vec{r}) \left[ \left< \gamma_{nb}^{\dagger}\gamma_{mb}^{}
\right> -\delta_{mn}\right],
\end{align}
and for the ground-state gap
\begin{equation}\label{eq:Delta0}
\Delta_{\text{GS}}(\vec{r}) =  g \sum_{m} u_m(\vec{r}) v_m^*(\vec{r}).
\end{equation} 
Therefore, a diagonalization of Eq. \eqref{eq:Matrix} requires a self-consistent treatment together with Eq. \eqref{eq:Delta0}. In doing so, we set the chemical potential $\mu=E_F$ with $E_F$ the Fermi energy of the bare atomic system, i.e., we assume that the chemical potential is not effected by the BCS pairing. Strictly speaking this assumption is only valid in the deep BCS regime. However, our numerical data on the basis of the Anderson approximation (see section \ref{ssec:anderson}) show that it has no qualitative effect on the features studied in this work.

A final remark to Eq. \eqref{eq:Delta0}: In the presented form the gap equation exhibits an ultraviolet-divergence, i.e., --strictly speaking-- Eq. \eqref{eq:Delta0} needs to be regularized to ensure the convergence of the sum over the BdG eigenstates \cite{Bloch2008Many}. However, in our case we restrict those sums to a rather narrow energy range around the Fermi level (see above) and this numerical cutoff remedies the need for a further regularization. This will be different in the calculations based on the Anderson approximation, as will be discussed below.  

\subsection{Quench dynamics}\label{ssec:dynamics}

To calculate the dynamics of the BCS condensate we make use of the Bogoliubov transformation Eqs. \eqref{eq:B-Transformation1} and \eqref{eq:B-Transformation2}. The quasiparticles resulting from that transformation are the single-particle excitations of the BCS phase, which are created, when the system is perturbed. All dynamical quantities investigated in the context of this work can be expressed in terms of expectation values of these quasiparticles. Therefore, we use Heisenberg's equation of motion for the quasiparticle operators to numerically calculate the dynamics of the quasiparticle expectation values.

To this end, we express the BdG Hamiltonian in terms of the single-particle operators for a general nonequilibrium situation where $\Delta(\vec{r},t) \neq \Delta_{\text{GS}}(\vec{r})$, i.e., where the current value of the gap differs from the ground-state value $\Delta_{\text{GS}}(\vec{r})$. Thus, inserting Eqs. \eqref{eq:B-Transformation1} and \eqref{eq:B-Transformation2} into Eq. \eqref{eq:BDG} where $\Delta = \Delta(\vec{r},t) \neq \Delta_{\text{GS}}(\vec{r})$ and identifiying $\Delta_{\text{GS}}$ via Eq. \eqref{eq:Delta0} yields
\begin{align}
	\lefteqn{H_{\mathrm{BdG}} = \sum_m E_{ma} \left(\gamma_{ma}^{\dagger} \gamma_{ma}^{}
		+ \gamma_{mb}^{\dagger}\gamma_{mb}^{} - 1\right)} \notag \\
	&+ \sum_{m,n} \left[\left(\Delta-\Delta_{\text{GS}}\right)_{u_m^* v_n}
	+ \left(\Delta^*-\Delta_{\text{GS}}^*\right)_{v_m^* u_n} \right]
	\gamma_{ma}^{\dagger}\gamma_{na}^{} \notag \\
	&+ \sum_{m,n} \left[\left(\Delta-\Delta_{\text{GS}}\right)_{u_m^* u_n^*}
	- \left(\Delta^*-\Delta_{\text{GS}}^*\right)_{v_m^* v_n^*} \right]
	\gamma_{ma}^{\dagger}\gamma_{nb}^{\dagger} \notag
	\end{align}
	\begin{align}\label{eq:BdG-Hamilton}
	\hspace{-1cm} &- \sum_{m,n} \Big[\left(\Delta-\Delta_{\text{GS}}\right)_{v_m v_n} -
	\left(\Delta^*-\Delta_{\text{GS}}^*\right)_{u_m u_n} \Big]
	\gamma_{mb}^{}\gamma_{na}^{} \notag \\
	&+ \sum_{m,n}  \left[\left(\Delta-\Delta_{\text{GS}}\right)_{v_m u_n^*}
	+ \left(\Delta^*-\Delta_{\text{GS}}^*\right)_{u_m v_n^*} \right]
	\gamma_{mb}^{\dagger}\gamma_{nb}^{} \notag\\
    &- \sum_{m,n}  \left[\left(\Delta-\Delta_{\text{GS}}\right)_{v_m u_n^*}
	+ \left(\Delta^*-\Delta_{\text{GS}}^*\right)_{u_m v_n^*} \right], 
\end{align}
with 
\begin{align}
(\Delta-\Delta_{\text{GS}})_{u_m v_n} &:= \nonumber \\
 \int& \mathrm{d}^3 r u_m^*(\vec{r}) \big[\Delta(\vec{r},t)-\Delta_{\text{GS}}(\vec{r})\big] v_n(\vec{r}).
\end{align}
We insert this into Heisenberg's equation of motion
\begin{equation} 
\dfrac{d}{dt} A_H = \dfrac{i}{\hbar}\left[ H_H, A_H \right] + \left(\dfrac{\partial}{\partial t} A\right)_H,
\end{equation}
where $A_H$ is the corresponding operator in the Heisenberg picture and $\left(\dfrac{\partial}{\partial t} A\right)_H = 0$ for the quasiparticle operators since all our calculations are performed with fixed basis states $\left[u_m(\vec{r}), v_m(\vec{r})\right]$. For the required single-particle expectation values this yields the following equations of motion:
\begin{widetext}
\begin{align}\label{eq:BwGl1}
	i\hbar \frac{d}{dt} \langle \gamma_{ma}^\dagger & \gamma_{na}\rangle = - (E_m - E_n)\langle \gamma_{ma}^\dagger \gamma_{na}\rangle + \sum\limits_l\Big( \notag \\
	&- \left[\left(\Delta-\Delta_{\text{GS}}\right)_{u_l^* v_m} +	
	 \left(\Delta^*-\Delta_{\text{GS}}^*\right)_{v_l^* u_m} \right]
	\langle \gamma_{la}^{\dagger}\gamma_{na}^{}\rangle 
	+ \left[\left(\Delta-\Delta_{\text{GS}}\right)_{u_n^* v_l}
	+ \left(\Delta^*-\Delta_{\text{GS}}^*\right)_{v_n^* u_l} \right]
	\langle \gamma_{ma}^{\dagger}\gamma_{la}\rangle  \notag \\
	&+ \left[\left(\Delta-\Delta_{\text{GS}}\right)_{u_n^* u_l^*} -
	\left(\Delta^*-\Delta_{\text{GS}}^*\right)_{v_n^* v_l^*} \right]
	\langle\gamma_{ma}^{\dagger}\gamma_{lb}^{\dagger}\rangle 
	-  \left[-\left(\Delta-\Delta_{\text{GS}}\right)_{v_l v_m}
	+ \left(\Delta^*-\Delta_{\text{GS}}^*\right)_{u_l u_m} \right] \langle\gamma_{lb}^{}\gamma_{na}^{}\rangle\Big)
\end{align}
\begin{align}\label{eq:BwGl2}
	i\hbar \frac{d}{dt} \langle \gamma_{mb}^\dagger & \gamma_{nb}\rangle = - (E_m - E_n)\langle \gamma_{mb}^\dagger \gamma_{nb}\rangle + \sum\limits_l\Big( \notag \\
	&- \left[\left(\Delta-\Delta_{\text{GS}}\right)_{u_l^* v_m} +	
	 \left(\Delta^*-\Delta_{\text{GS}}^*\right)_{v_l^* u_m} \right]
	\langle \gamma_{lb}^{\dagger}\gamma_{nb}^{}\rangle 
	+ \left[\left(\Delta-\Delta_{\text{GS}}\right)_{u_n^* v_l}
	+ \left(\Delta^*-\Delta_{\text{GS}}^*\right)_{v_n^* u_l} \right]
	\langle \gamma_{mb}^{\dagger}\gamma_{lb}\rangle  \notag \\
	&+ \left[\left(\Delta-\Delta_{\text{GS}}\right)_{u_n^* u_l^*} -
	\left(\Delta^*-\Delta_{\text{GS}}^*\right)_{v_n^* v_l^*} \right]
	\langle\gamma_{la}^{\dagger}\gamma_{mb}^{\dagger}\rangle 
	-  \left[-\left(\Delta-\Delta_{\text{GS}}\right)_{v_l v_m}
	+ \left(\Delta^*-\Delta_{\text{GS}}^*\right)_{u_l u_m} \right] \langle\gamma_{nb}^{}\gamma_{la}^{}\rangle\Big)
\end{align}
\begin{align}\label{eq:BwGl3}
	i\hbar \frac{d}{dt} \langle \gamma_{mb} & \gamma_{na}\rangle = - i\hbar \frac{d}{dt} \Big( \langle \gamma_{na}^\dagger  \gamma_{mb}^\dagger\rangle \Big)^* = (E_m + E_n)\langle \gamma_{mb} \gamma_{na}\rangle + \sum\limits_l\Big( \notag \\
	&\quad \left[\left(\Delta-\Delta_{\text{GS}}\right)_{u_n^* v_l} +	
	 \left(\Delta^*-\Delta_{\text{GS}}^*\right)_{v_n^* u_l} \right]
	\langle \gamma_{mb} \gamma_{la}\rangle 
	+ \left[\left(\Delta-\Delta_{\text{GS}}\right)_{u_n^* u_l^*}
	- \left(\Delta^*-\Delta_{\text{GS}}^*\right)_{v_n^* v_l^*} \right]
	\left( \delta_{ml} - \langle \gamma_{lb}^{\dagger}\gamma_{mb}\rangle \right)  \notag \\
	&- \left[\left(\Delta-\Delta_{\text{GS}}\right)_{u_l^* u_m^*} -
	\left(\Delta^*-\Delta_{\text{GS}}^*\right)_{v_l^* v_m^*} \right]
	\langle\gamma_{la}^{\dagger}\gamma_{na}^{}\rangle 
	+  \left[-\left(\Delta-\Delta_{\text{GS}}\right)_{v_l u_m^*}
	+ \left(\Delta^*-\Delta_{\text{GS}}^*\right)_{u_l v_m^*} \right] \langle\gamma_{lb}^{}\gamma_{na}^{}\rangle\Big).
\end{align}
\end{widetext}
We solve these nonlinearly coupled equations of motion for the initial value problem defined by an instantaneous interaction quench. I.e., we start from the ground state corresponding to a scattering length $a_i$ and instantaneously switch to a different value\footnote{This assumption of an instantaneous quench is valid, since the experimentally achieved quench times $\sim$ns are way below the timescales of the gap dynamics $\sim$ms.} $a_i \rightarrow a_f$. Therefore, during the quench at the time $t=0$ the system has no time to relax to the new ground state corresponding to $a_f$ but it remains in the old ground state corresponding to $a_i$. I.e., all quasiparticle expectation values of the form $\langle \gamma_{M}^{\dagger} \gamma_{M^\prime} \rangle$ and $\langle \gamma_{M} \gamma_{M^\prime} \rangle$ with $M \in \{ma, mb\}$ and $M^{\prime} \in \{m^{\prime}a, m^{\prime}b\}$ which correspond to the old ground state vanish for $t=0$. With this in mind, we invert the Bogoliubov transformation for the operators $\gamma_{M}$ in the old basis before the quench and insert the resulting expressions for $\Psi_\sigma^{ }$ and $\Psi_\sigma^{\dagger}$ into the quasiparticle operators in the new basis after the quench [Eqs. \eqref{eq:B-Transformation1} and \eqref{eq:B-Transformation2}]. This yields the initial values for the dynamics
\begin{align}\label{eq:Start1}
	\langle \gamma_{ma}^\dagger  & \gamma_{na}\rangle|_{t=0} = 	\sum\limits_k \int \mathrm{d}^3 r \Big[ v_m(\vec{r}) \tilde{u}_k(\vec{r}) -u_m(\vec{r})\tilde{v}_k(\vec{r}) \Big] \notag \\
	 &\qquad\cdot\int \mathrm{d}^3 r^\prime \Big[v_n^*(\vec{r}^\prime) \tilde{u}_k^*(\vec{r}^\prime)-u_n^*(\vec{r}^\prime) \tilde{v}_k^*(\vec{r}^\prime) \Big]
\end{align}
\begin{align}\label{eq:Start2}
\langle \gamma_{ma}^\dagger  & \gamma_{nb}^\dagger\rangle|_{t=0} = 	\sum\limits_k \int \mathrm{d}^3 r \Big[ v_m(\vec{r}) \tilde{u}_k(\vec{r}) - u_m(\vec{r})\tilde{v}_k(\vec{r}) \Big] \notag \\
	 &\qquad\cdot\int \mathrm{d}^3 r^\prime \Big[v_n(\vec{r}^\prime) \tilde{v}_k^*(\vec{r}^\prime)+u_n(\vec{r}^\prime) \tilde{u}_k^*(\vec{r}^\prime) \Big]\\
\langle \gamma_{mb} & \gamma_{na} \rangle|_{t=0} = \Big( \langle \gamma_{na}^\dagger  \gamma_{mb}^\dagger\rangle|_{t=0} \Big)^*
\end{align}
\begin{align}\label{eq:Start3}
\langle \gamma_{mb}^\dagger & \gamma_{nb}\rangle|_{t=0} = - \sum\limits_k \int \mathrm{d}^3 r \Big[ v_m^*(\vec{r}) \tilde{v}_k(\vec{r}) + u_m^*(\vec{r})\tilde{u}_k(\vec{r}) \Big] \notag \\
	 &\cdot \int \mathrm{d}^3 r^\prime \Big[v_n(\vec{r}^\prime) \tilde{v}_k^*(\vec{r}^\prime)+u_n(\vec{r}^\prime) \tilde{u}_k^*(\vec{r}^\prime) \Big] + \delta_{mn},
\end{align}
where $\tilde{u}_m$ and $\tilde{v}_m$ refer to the single-particle states before the quench and $u_m$ and $v_m$ to those after the quench.\\

With Eqs. \eqref{eq:Start1}-\eqref{eq:Start3} we can numerically integrate the equations of motion and thus calculate the gap dynamics for a gas of $^6$Li in a 3D harmonic trap via Eq. \eqref{eq:Delta_Bogolon}. In section \ref{ssec:gap} we will present the corresponding results for the phase dynamics of the gap. However, before that we introduce Anderson's approximation which we will use to perform elaborate parameter scans and to calculate the gap dynamics for systems with rather large particle number that are numerically too complex to address within the full BdG approach. In doing so, we will restrict our explanations to the main aspects of the approximation. A detailed description of the corresponding formalism with all expressions derived from the above can be found in our previous work \cite{hannibal2015quench}.

\subsection{Anderson's Approximation}\label{ssec:anderson}

In Anderson approximation the expansion of the quasiparticle wave function in Eqs. \eqref{eq:entw1} and \eqref{eq:entw2} is truncated such that $u_m(\vec{r})=u_m \phi_m(\vec{r})$ and $v_m(\vec{r})=v_m \phi_m(\vec{r})$. This strongly simplifies the formalism presented above. For the ground-state properties the diagonalization of Eq. \eqref{eq:Matrix} directly yields
\begin{equation}\label{eq:Eq}
	E_m = \sqrt{(\varepsilon_m-\mu)^2 + (\Delta^{\text{GS}}_{mm})^2}
\end{equation}
and
\begin{equation}\label{eq:uv}
		u_m = \sqrt{\frac{1}{2} \left(1+ \frac{\varepsilon_m-\mu}{E_m}\right)} \qquad  v_m = \sqrt{\frac{1}{2} \left(1 - \frac{\varepsilon_m-\mu}{E_m}\right)}.
\end{equation} 
which --in combination with Eq. \eqref{eq:Delta0}-- leads to a BCS-like selfconsistency equation, that we solve numerically. In doing so, we use much larger energy windows in the sum over the states as compared to the full BdG approach since the numerical effort is strongly reduced in the case of the Anderson approximation. Thereby, we employ the regularization scheme introduced in \cite{hannibal2015quench}. Only when directly comparing the full BdG equations with the Anderson approximation (Sec. \ref{ssec:gap}) we use the same cutoffs in both calculations to improve the comparability of the respective calculations.

For the dynamical situation we furthermore assume that $\left(\Delta-\Delta_{\text{GS}}\right)_{x_m y_n} = \left(\Delta-\Delta_{\text{GS}}\right)_{x_m y_n} \delta_{mn}$ with $x_m,y_m \in \{u_m, v_m\}$, i.e., that the main implications of the Anderson approximation $\Delta^{\text{GS}}_{mn}=\Delta^{\text{GS}}_{mn} \delta_{mn}$ holds for the dynamical situation as well. That leads to a great simplification of the equations of motion.

However, strictly speaking Anderson's approximation is only valid if $\Delta^{\text{GS}}_{mn} \ll \delta \varepsilon$ with $\delta \varepsilon$ the level spacing of the harmonic eigenenergies. In general, this only holds for weak coupling, i.e., deep in the BCS regime, and/or for strong confinements and thus large level spacing. Nevertheless, our numerical data show that the approximation reproduces all the main features investigated in this work even for moderate confinements in the BCS-BEC crossover regime (cf. section \ref{ssec:gap}).

\section{Results}\label{sec:results}

In the following, we investigate the phase dynamics of the spatially averaged BCS gap\footnote{In our case the phase of the gap $\varphi$ is nearly homogeneous (see below), i.e., $\bar{\varphi} \approx \varphi(\vec{r})$.}
\begin{equation}
\bar{\Delta}(t) = \frac{1}{V}\int \mathrm{d}^3r \Delta(\vec{r},t)
\end{equation}
for an ultracold gas of $^6$Li in a cigar-shaped harmonic trap. We use $V=l_x l_y l_z$ with $l_\alpha$ being the oscillator length in direction $\alpha$ as a normalization volume. We excite the system by interaction quenches $a_i \rightarrow a_f$.

In section \ref{ssec:gap} we will identify the emergence of a Goldstone mode in the phase dynamics of the BCS gap with one dominant low-frequency contribution. Furthermore, we will show that the results obtained within the Anderson approximation are in good agreement with the full BdG solution. In section \ref{ssec:g-scan} we will analyze this Goldstone mode over a wide range of parameters and we will show that its excitation spectrum is gapless and that its main frequency is not determined by the trap parameters but by the details of the excitation. 

In section \ref{ssec:resonances} we will investigate the influence of the confinement parameters on the phase dynamics and by that the effect of the superfluid resonances found in \cite{Shanenko2012Atypical}. Furthermore, in section \ref{ssec:single} we will evaluate the impact of the interaction quench-induced Goldstone mode on the single-particle excitations of the cloud which could provide an experimental access to the gapless phase mode.   

\subsection{Phase dynamics of the gap}\label{ssec:gap}

\begin{figure}[t]
	\centering
	\includegraphics[width=1\columnwidth]{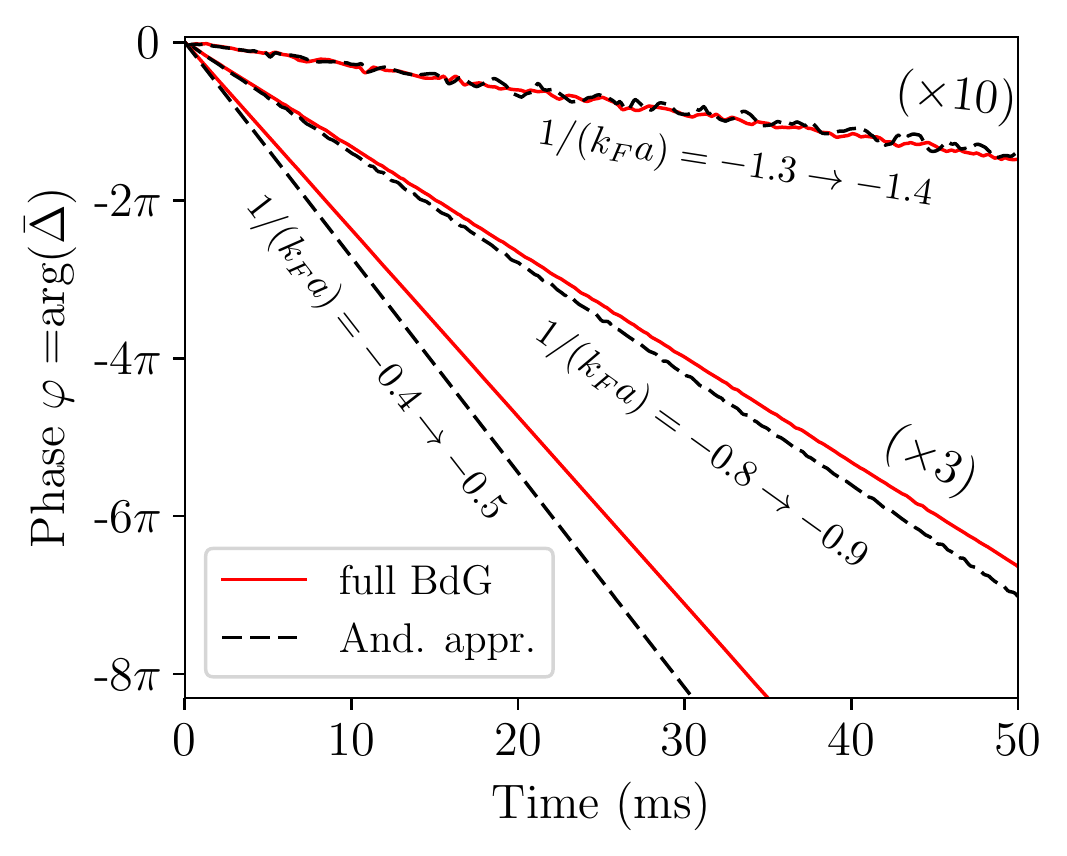}
	\caption{(color online) Phase dynamics for the full BdG solution (solid lines) and the Anderson approximate solution (dashed lines) for $\delta\left[1/(k_Fa)\right] = -0.1$ and different final coupling strengths: $1/(k_Fa_f)=-0.5$, $1/(k_Fa_f)=-0.9$ (scaled by a factor of 3) and $1/(k_Fa_f)=-1.4$ (scaled by a factor of 10); parameters: $f_\parallel = 96\,$Hz, $f_\perp = 1\,$kHz, $N_P=120$.}
	\label{fig:GM}
\end{figure}

Figure \ref{fig:GM} shows the dynamics of the phase $\varphi=\text{arg}(\bar{\Delta})$ of the spatially averaged BCS gap for a system with $f_\perp = 1\,\mathrm{kHz}$ and $f_\parallel = 96\,\mathrm{Hz}$ excited by quenches with the strength $\delta\left[1/(k_Fa)\right] = 1/(k_Fa_f) - 1/(k_Fa_i) = -0.1$ at different positions in the BCS-BEC crossover. The particle number is set to $N_P=120$ and the expansion of the single-particle wave functions in Eqs. \eqref{eq:entw1} and \eqref{eq:entw2} is restricted to atomic states from a window of width $\Delta\epsilon:=0.92E_F$ around the chemical potential (i.e., $\mu - 0.46 E_F \leq \varepsilon \leq \mu + 0.46 E_F$). This is the limitation of our current numerical setup for the full BdG approach.

The solid lines in Fig. \ref{fig:GM} show the data obtained by the full equations of motion \eqref{eq:BwGl1} - \eqref{eq:BwGl3}.
First of all, one clearly observes that --for all quenches-- the phase dynamics of the gap is strongly dominated by a linear decrease in time. Therefore, after the quench the system performes a constant phase \enquote{motion} of the gap which nicely corresponds to the simplified picture of a Mexican-hat potential introduced in Fig. \ref{fig:Mexican}: The potential is flat inside the rim which implies a constant phase velocity, i.e., a steady oscillation inside the rim where the frequency of the latter --i.e., the frequency of the Goldstone mode-- is defined by the time-averaged slope $f_G := \frac{1}{2\pi} \frac{\Delta\varphi}{\Delta t}$ where $\Delta t$ is large compared to the intrinsic time scales of the system.\footnote{We want to remark that the appearance of a Goldstone mode with a fixed frequency is similar to the case of the AC Josephson effect. There, a difference in the chemical potentials $\mu_i$ on the two sides of a Josephson junction leads to the emergence of a Goldstone mode with the frequency given by $f_G \sim (\mu_2 - \mu_1)$ \cite{pekker2015amplitude}. In contrast, in our case the Goldstone mode is driven by the quench which drives the system instantaneously from an equilibrium into a non-equilibrium state.} The corresponding values are given by $f_G=1.6\,$Hz for the weakest coupling strength and $f_G=21.9\,$Hz for $1/(k_Fa_f)=-0.9$ and $f_G=118.7\,$Hz for $1/(k_Fa_f)=-0.5$ (for illustrative purposes the curves for the two weaker coupling strengths are scaled by a factor 3 and 10, respectively). Thus, the interaction quench-induced \enquote{phase velocity} strongly increases when approaching the unitary point $1/(k_Fa) = 0$.

Furthermore, a closer look at Fig. \ref{fig:GM} reveals that a higher-frequency oscillation exists on top of the linear contribution. This contribution is strongest for the system with $1/(k_Fa_f)=-1.4$ and much weaker --and thus not directly visible in Fig. \ref{fig:GM}-- for the stronger-coupling cases. A more detailed analysis shows that the corresponding frequencies again increase when approaching the unitary point, i.e., this dynamics is fast for large and slow for small coupling strengths. Nevertheless, the corresponding range of frequencies coincides with that from the spectrum of the Higgs mode. This indicates that the Higgs and the Goldstone mode are weakly coupled, where we observe that the influence of the Higgs mode increases when approaching the BCS limit $1/(k_Fa) \ll -1$ and when increasing the modulus of quench strength $|\delta\left[1/(k_Fa)\right]|$. However, the Higgs mode was extensively studied in \cite{hannibal2015quench}. Therefore, in this work we will not go into details about these contributions. 

The dashed lines in Fig. \ref{fig:GM} show the Anderson approximate solution corresponding to section \ref{ssec:anderson} (note that in order to improve the comparability here we have used an ungregularized Anderson solution with the same cutoff as in the calculations without Anderson approximation). One can see that the approximate solution gives an overall good qualitative agreement with the full dynamics: It shows the same linear decrease in time with a --in the stronger coupling cases not directly visible-- higher-frequency contribution on top. Again, the slow linear decrease corresponds to the Goldstone mode while the higher-frequency component results from the coupling to the Higgs mode\footnote{The agreement of both solutions with respect to the Higgs mode may not look very convincing. However, already rather small deviations in the spectral composition of the Higgs mode result in large deviations in the time domain at larger times. I.e., the qualitative agreement of both solutions is in indeed good.}. However, one can observe as well that the quantitative deviations between the full and the approximate solution increase with increasing coupling strength: While --considering that the corresponding curves in Fig. \ref{fig:GM} are scaled by a factor of 10-- the frequency of the Goldstone mode, i.e., the slope of the linear contribution, matches very well for the case of $1/(k_Fa_f) = -1.4$  the deviation of the two solutions becomes rather significant when increasing the coupling strength.

\begin{figure}[t]
	\centering
	\includegraphics[width=0.9\columnwidth]{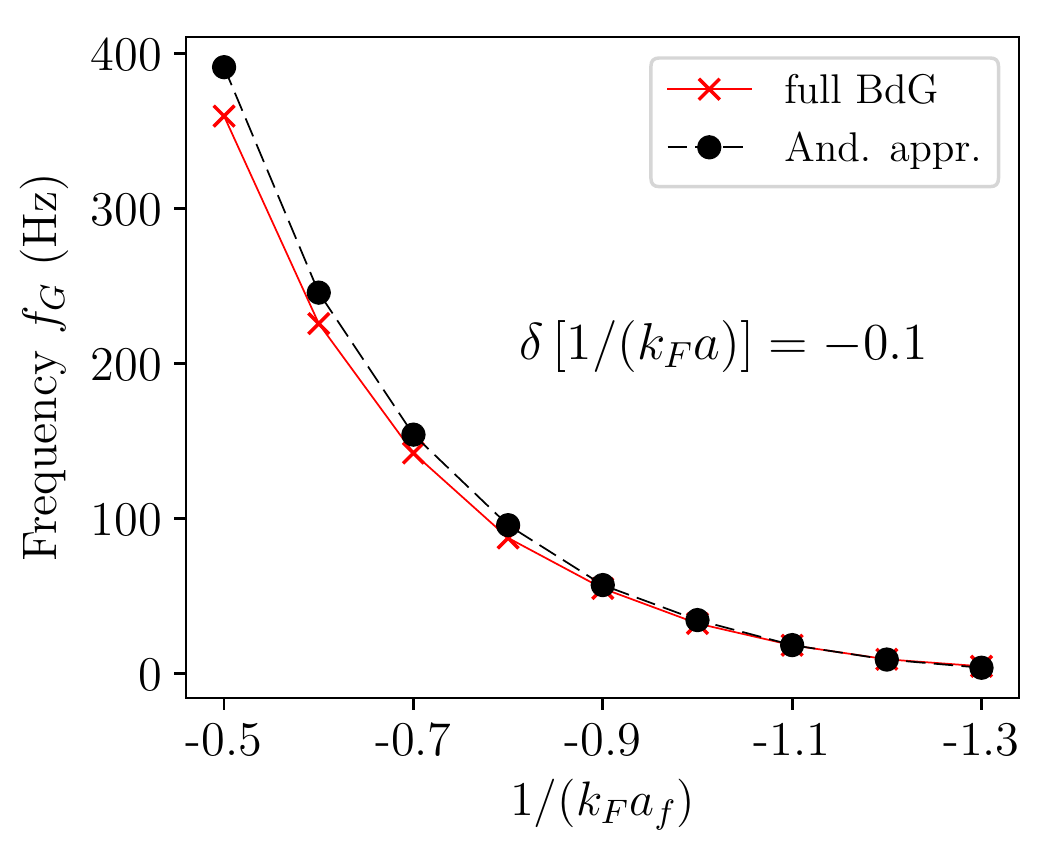}
	\caption{(color online) Frequency of the Goldstone mode resulting from the full BdG solution (solid line) and from Anderson's approximate solution (dashed line) for the same system as in Fig. \ref{fig:GM} but varying $1/(k_Fa_f)$ (to reduce the numerical effort we set $\Delta\epsilon = 0.58E_F$).}
	\label{fig:fG}
\end{figure}

This is illustrated in Fig. \ref{fig:fG}. There, the frequencies of the Goldstone mode resulting from the full and from the approximate solution are shown for a fixed quench strength $\delta\left[1/(k_Fa)\right] = -0.1$ and varying $1/(k_Fa_f)$. One can clearly observe that the approximate solution reproduces well the qualitative trend, i.e., an overall increase of the frequency of the Goldstone mode when approaching the unitary point. Indeed, when entering the BCS regime even the quantitative values match very well. However, one can see as well, that the deviation between both solutions increases with increasing coupling strength and becomes rather large in the crossover regime. This indicates that Anderson's approximation tends to break down --as expected-- when approaching the unitary point.    

Nevertheless, we want to emphasize that all features in the phase dynamics investigated in this work are fully reproduced by the Anderson approximation. Solely the shift in frequency between the full and the approximate solution increases with increasing coupling strength. Therefore, all following calculations of this work are performed in Anderson approximation which allows for a drastical reduction of computational effort and thus for a detailed investigation of parameter dependencies and of larger systems at all. In doing so, from now on all sums will be restricted to states from a window of size $2E_F$ around the Fermi level instead of $0.92E_F$ as before to ensure a better quantitative convergence of the obtained frequencies (cf. \cite{hannibal2015quench}). 

Continuing in Anderson approximation we will now investigate the phase dynamics in closer detail, i.e., we will analyze the influence of the external parameters on the frequency $f_G$ of the Goldstone mode and we will point out by which quantities it is determined. To do so, we will first investigate the influence of the coupling strength and of the details of the quench. In section \ref{ssec:resonances} we will focus on the effect of the confinement.

\subsection{Influence of the quench}\label{ssec:g-scan}

\begin{figure}[t]
	\centering
	\includegraphics[width=0.9\columnwidth]{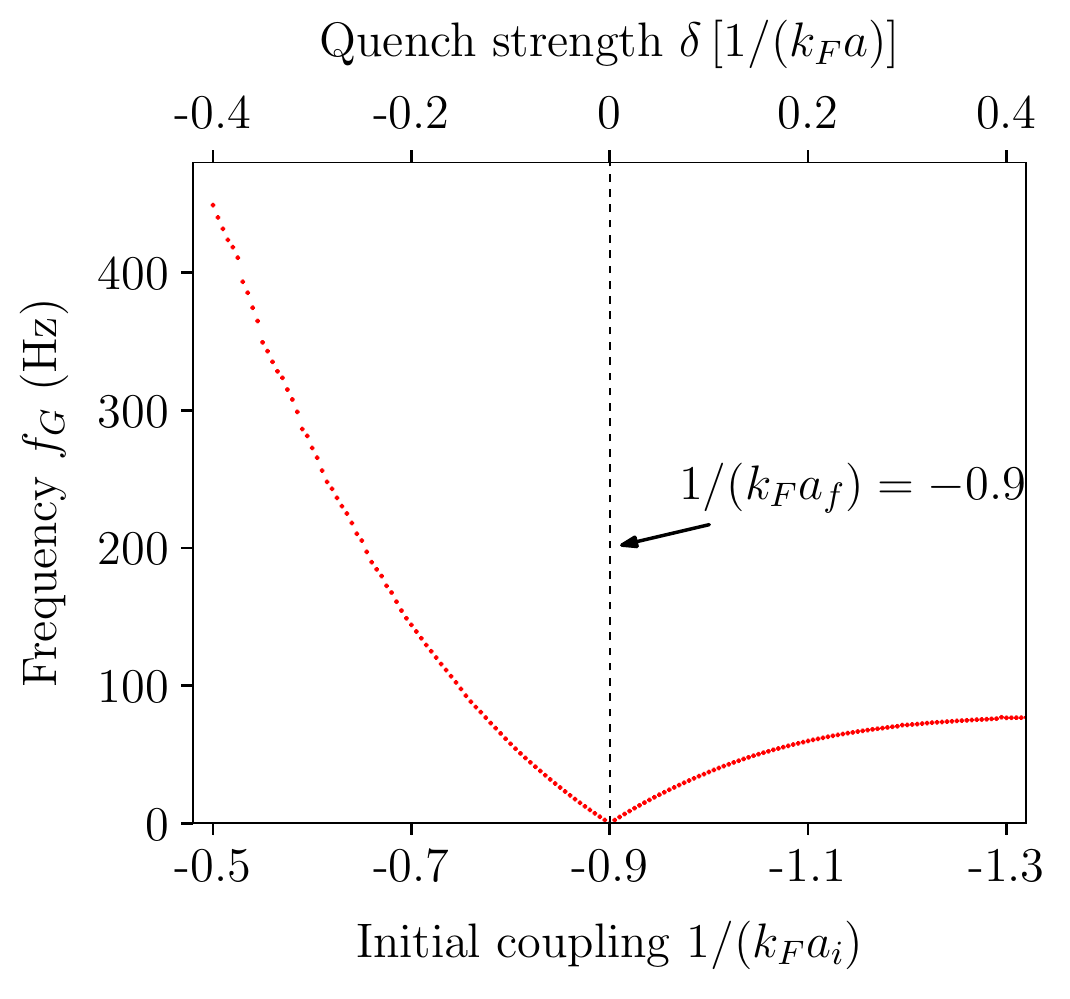}
	\caption{(color online) Frequency of the Goldstone mode $f_G$ for quenches with varying strength to $1/(k_Fa_f)=-0.9$; parameters: $f_\parallel = 96\,$Hz, $f_\perp = 1\,$kHz, $N=1000$.}
	\label{fig:strength}
\end{figure}

\begin{figure}[t]
	\centering
	\includegraphics[width=1\columnwidth]{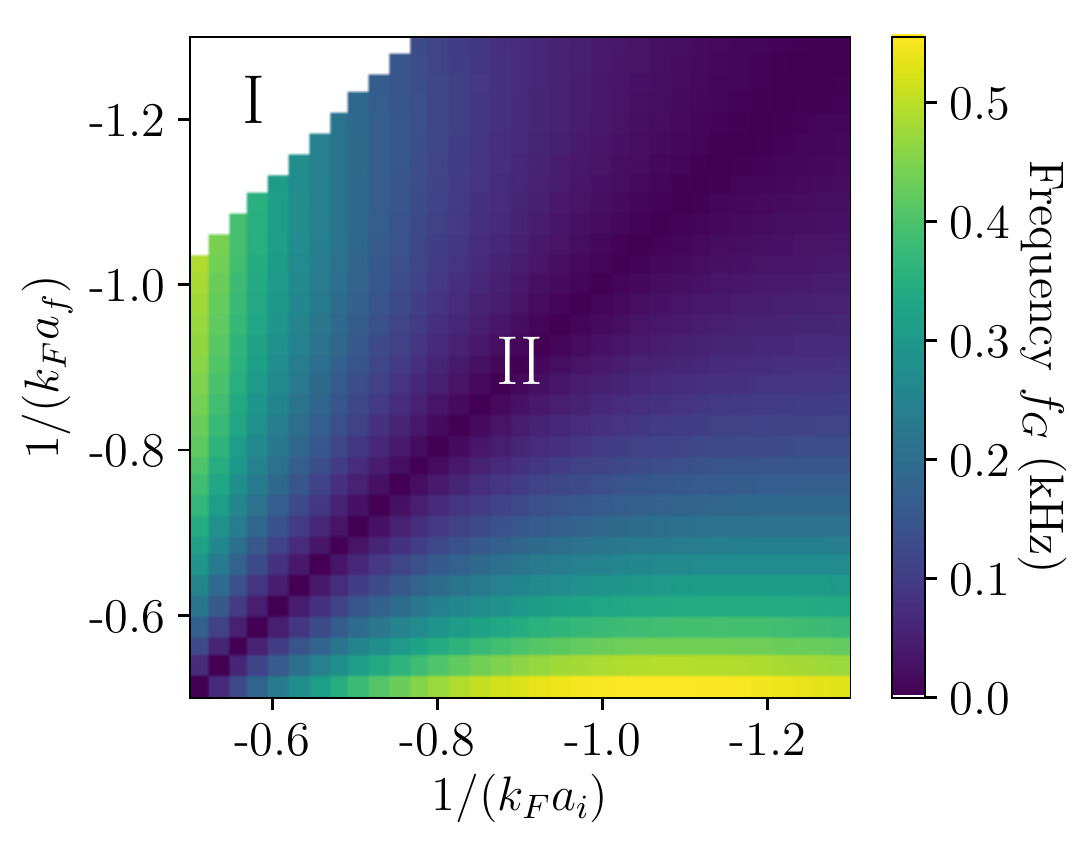}
	\caption{(color online) Frequency of the Goldstone mode $f_G$ for different excitations $1/(k_Fa_i) \rightarrow 1/(k_Fa_f)$ corresponding to phase II of the quantum quench phase diagram (see main text); quenches corresponding to phase I are not shown; parameters: $f_\parallel = 96\,$Hz, $f_\perp = 1\,$kHz, $N=1000$. }
	\label{fig:g_scan}
\end{figure}

Figure \ref{fig:GM} already suggests that the frequency of the Goldstone mode $f_G$ depends on the details of the quench instead of being fixed by the external parameters of the cloud. The phase dynamics changes from a very slow decrease for the system in the BCS regime with $1/(k_Fa_f) = -1.4$ to a rather fast decrease in the crossover regime with $1/(k_Fa_f) = -0.5$. However, in Fig. \ref{fig:GM} the quench strength by means of $\delta\left[1/(k_Fa)\right]$ was kept fixed. Only the position in the BCS-BEC crossover was varied. Accordingly, we will in the following investigate the influence of the quench strength on the phase dynamics for one particular final coupling strength $1/(k_Fa_f)$ in Fig. \ref{fig:strength} and for a wider range of $1/(k_Fa_f)$ in Fig. \ref{fig:g_scan}. In doing so, we will show, that the excitation spectrum of the interaction quench-induced Goldstone mode is gapless, i.e., it is a Goldstone mode in the original sense of the Goldstone theorem. Furthermore, we will show that its frequency can be adjusted by changing the details of the quench and that it is determined by the initial values of the dynamics and by the gap of the system after the quench. 

Figure \ref{fig:strength} shows the frequency of the Goldstone mode for a system with $1/(k_Fa_f)=-0.9$ in the same trap as in Fig. \ref{fig:GM} but --since we now apply Anderson's approximation-- for a larger particle number\footnote{With our current numerical setup we can calculate the dynamics of single systems for up to $N_P \sim 10^4$ particles. However, there are no qualitative changes in the gap dynamics when increasing the particle number, i.e., here we restrict ourselves to rather small $N_P$ to reduce the numerical effort.} of $N_P = 1000$. The initial coupling strength $1/(k_Fa_i)$ is varied, i.e., the dependence of $f_G$ on the quench strength is shown. 

First of all, Fig. \ref{fig:strength} demonstrates that the phase dynamics strongly depends on the quench strength: When approaching $1/(k_Fa_i) = 1/(k_Fa_f) = -0.9$, i.e., the point of quench strength $\delta\left[1/(k_Fa)\right]=0$, from either side [from larger or smaller values of $1/(k_Fa_i)$] the frequency of the Goldstone mode continuously decreases to zero. Thus, the frequency of the Goldstone mode decreases with decreasing (modulus of the) quench strength and continuously vanishes for $\delta\left[1/(k_Fa)\right]\rightarrow0$. This implies that the excitation spectrum of the interaction quench-induced phase mode is indeed gapless as stated above.

Furthermore, we observe that the dependence of $f_G$ on the quench strength in Fig. \ref{fig:strength} is asymmetric: Negative quenches with $\delta\left[1/(k_Fa)\right]<0$, i.e., those on the left hand side of Fig. \ref{fig:strength}, lead to a stronger increase in $f_G$ and thus to larger frequencies than positive quenches. This asymmetry is linked to the fact that the same excitation strength in terms of $1/(k_Fa)$, i.e., the same $|\delta\left[1/(k_Fa)\right]|$, results in different actual changes in the scattering length $|\delta a| = |a_f - a_i|$ depending on the position of the initial system  in the crossover. Hence, negative quenches lead to much larger changes in the scattering length $|\delta a|$ --and therefore in the gap-- than positive quenches. As we will show later on, this results in larger $f_G$. However, before we do so, we will demonstrate that the features found above hold for a wide range of quenches in the BCS-BEC crossover. Indeed, our numerical data indicate that the above found nature of the phase dynamics holds for all moderate quenches on the BCS side of the BCS-BEC crossover, i.e., for all quenches that can be associated with the phase II of the quantum quench phase diagram introduced in \cite{Yuzbashyan2015Quantum}. For quenches exceeding this range, the phase dynamics tends to become irregular. In particular, strong negative quenches which lead to a dynamical vanishing of the gap in the Higgs mode (phase I) exhibit a persistent but very irregular phase dynamics which makes the definition of a frequency of the Goldstone mode arbitrary. However, in this work we restrict ourselves to the investigation of the regular phase dynamics in the gapless Golstone mode, i.e., to quenches in the phase II.

To this end, Fig. \ref{fig:g_scan} shows the dependence of $f_G$ on the quench strength for varying $1/(k_Fa_f)$, i.e., each horizontal line in Fig. \ref{fig:g_scan} corresponds to a plot like in Fig. \ref{fig:strength} but for a different coupling strength.

Indeed, one observes on the basis of Fig. \ref{fig:g_scan} that the nature of the Goldstone mode described above holds for all phase II quenches investigated here (we omit the quenches in the upper left corner corresponding to phase I). For each horizontal line in Fig. \ref{fig:g_scan} we see a decrease of $f_G$ with decreasing modulus of the quench strength $|\delta\left[1/(k_Fa)\right]|$ as found above with a continously vanishing Goldstone mode when approaching $1/(k_Fa_i) = 1/(k_Fa_f)$. However, furthermore one can see, that the frequency of the Goldstone mode also depends on the vertical position of the quench in Fig. \ref{fig:g_scan}: It is largest for quenches at the bottom, i.e., for systems with large coupling strength. In fact, one can observe that quenches with the same $|\delta\left[1/(k_Fa)\right]|$ but opposite directions, e.g., those correspoding to the points $\left[1/(k_Fa_i), 1/(k_Fa_f) \right] = (-1.0,-0.5)$ and $\left[1/(k_Fa_i), 1/(k_Fa_f) \right] = (-0.5,-1.0)$ lead to significantly different $f_G$. This indicates that $f_G$ depends not only on the quench strength but also on the coupling strength of the system after the quench, i.e., a stronger coupling results in a larger $f_G$. In the following, we will explain both these features --the dependence on the quench strength and the dependence on the final coupling strength-- by taking into account that the only parameters of the dynamics affected by the quench are the initial values of the excitation and the gap after the quench. We will isolate the effects of both quantities on the basis of Fig. \ref{fig:g_scan} and Eqs. \eqref{eq:Start1}-\eqref{eq:Start3}.

To do so, we at first want to state that the frequency of the Goldstone mode depends linearly on the initial values of the dynamics which in Anderson approximation read (cf. \cite{hannibal2015quench} and Eqs. \eqref{eq:Start1}-\eqref{eq:Start3}) 
\begin{align}\label{eq:start_And}
	\Big<\gamma_{ma}^{\dagger}\gamma_{ma}^{}\Big>\big|_{t=0}
	&= \left(v_m \tilde{u}_m - u_m  \tilde{v}_m\right)^2 =: x_m^{(0)} \notag \\
	\left<\gamma_{ma}^{\dagger}\gamma_{mb}^{\dagger}\right>\big|_{t=0}
	&= \left(v_m \tilde{u}_m - u_m \tilde{v}_m \right) \left(v_m \tilde{v}_m + u_m \tilde{u}_m\right) \notag \\
	&=: y_m^{(0)},
\end{align}
with $\tilde{u}_m$ and $\tilde{v}_m$ ($u_m$ and $v_m$) being the Anderson amplitudes before (after) the quench [see Eq. \eqref{eq:uv}]. Indeed, our numerical data show that by artificially multiplying these initial values by a factor $k$ the frequency $f_G$ increases by the same factor\footnote{For the situation of strong interactions and strong quenches even an analytical expression can be found: $\omega_G = 2\pi f_G  \approx \mathrm{Im}(\frac{d}{dt}\bar{\Delta})|_{t=0}/\bar{\Delta}_{\text{GS}}$.}, i.e.,    
\begin{equation}
f_G \left(k \cdot \{ x_m^{(0)}, y_m^{(0)}\}\right) = k f_G\left(\{ x_m^{(0)}, y_m^{(0)}\}\right). \nonumber
\end{equation}
This means that --whatever the actual system parameters ($a_f$, $N_P$, $f_\parallel$, $f_\perp$) are-- the frequency of the Goldstone mode can be tuned by adjusting the initial values of the dynamics. As Eqs. \eqref{eq:uv} and \eqref{eq:start_And} imply, the latter depend on the gaps of the system before and after the quench which are defined by the quench.

A detailed evaluation of Eqs. \eqref{eq:start_And} shows that the initial values are large when the difference between the gaps before and after the quench and thus between the amplitudes $(\tilde{u}_m, \tilde{v}_m)$ and $(u_m, v_m)$ is large. This is the case for strong quenches, i.e., strong quenches result in large initial values which well coincides with the above findings for $f_G$. However, Eqs. \eqref{eq:start_And} also show that the amplitude of the initial excitation is independent from the quench direction. When changing the quench direction, i.e., when interchanging $(\tilde{u}_m, \tilde{v}_m) \Leftrightarrow (u_m, v_m)$ in Eqs. \eqref{eq:start_And}, only the sign of the anomalous initial values $\left<\gamma_{ma}^{\dagger}\gamma_{mb}^{\dagger}\right>\big|_{t=0}$ changes. Therefore, the initial values for quenches with the same $|\delta\left[1/(k_Fa)\right]|$ but opposite directions have the same strength but --as observed above-- the frequency is larger for the respective positive quench to the stronger coupling $1/(k_Fa_f)$. This indicates that the frequency of the Goldstone mode also depends on the actual coupling strength of the system after the quench.

We can conclude the results of this paragraph: The excitation spectrum of the Goldstone mode of the BCS gap of a 3D confined ultracold Fermi gas excited by an interaction quench is gapless and the frequency of the Goldstone mode can be tuned in a wide range by adjusting the strength of the quench. Again, this coincides with the simplified picture of Fig. \ref{fig:Mexican}: However small the quench-induced initial \enquote{momentum} of the phase dynamics might be, it results in a constant \enquote{phase motion} inside the rim. This is --in simple words-- the consequence of the Goldstone theorem, which therefore can well be observed in the dynamics after an interaction quench.

However, the fact that in our case the Goldstone mode is gapless and that its frequency can be adjusted by the strength of the excitation stands in contrast to the experimental and theoretical findings of Refs. \cite{Kinast2004Evidence, Kinast2004Breakdown, bartenstein2004Collective, altmeyer2007dynamics,  altmeyer2007precision, riedl2008collective, baranov2000low, bruun2001low, bruun2002low, hu2004collective, heiselberg2004collective, Stringari2004Collective, grasso2005temperature, korolyuk2011density}. There, the frequency of the phase dynamics was found to be fixed by the frequencies of the trapping potential. Nevertheless, this discrepancy can be explained by the circumstance that in the previous works the dynamics was induced by spatially inhomogeneous perturbations of the cloud, e.g., by confinement quenches or optical excitations. Such a spatial perturbation creates a motion of the superfluid in the trap with a time-dependent verlocity $v_s$. This directly induces a dynamics of the phase of the gap via \cite{Giorgini2008Theory}
\begin{equation}
\vec{v_s}=\frac{\hbar}{2m} \vec{\nabla} \text{arg}\big[\Delta(\vec{r},t)\big].
\end{equation}
On the one hand, this means that --whenever a superfluid velocity $v_s$ is excited-- the real-space dynamics is directly linked to a dynamics in the Goldstone mode. In this sense, the latter can be observed through the motion of the cloud. On the other hand, the trapping potential governs the real-space dynamics of the cloud. Therefore, a direct coupling of the Goldstone mode to the real-space dynamics implies that the trap imprints its frequencies on the Goldstone mode. For inhomogeneous excitations $f_G$ is thus pushed to the trap frequencies as found in the previous works.\footnote{Actually, both manifestations of the Goldstone mode --the gapped inhomogeneous and the gapless one-- may exist at the same time depending on the nature of the excitation.}

In our case the excitation is spatially homogeneous and does not --as our numerical data confirm-- produce any significant phase gradients. Therefore, no real-space dynamics of the cloud is induced by the interaction quench which implies that the phase dynamics does not couple to the trap. Accordingly, the excitation spectrum of this \textit{homogeneous} Goldstone mode remains gapless. In this sense the interaction quench-induced Goldstone mode remains pure. 

\subsection{Impact of the superfluid resonances}\label{ssec:resonances}
In this section we want to study the influence of the confinement on the frequency of the gapless Goldstone mode and by that the impact of the size-dependend superfluid resonances theoretically predicted in \cite{Shanenko2012Atypical}. To do so, we will investigate the dependence of the phase dynamics on the trapping frequency in x-y direction $f_\perp$ for a system with fixed $f_\parallel = 96\,$Hz, $N_P = 1000$ atoms in the trap and a quench given by $1/(k_Fa) = -0.8 \rightarrow -0.9$. The complementary situation, i.e., a fixed $f_\perp$ with varying $f_\parallel$ produces the same effects. Therefore, the influence of $f_\parallel$ will not be investigated separately. 

\begin{figure}[t]
	\centering
	\includegraphics[width=1\columnwidth]{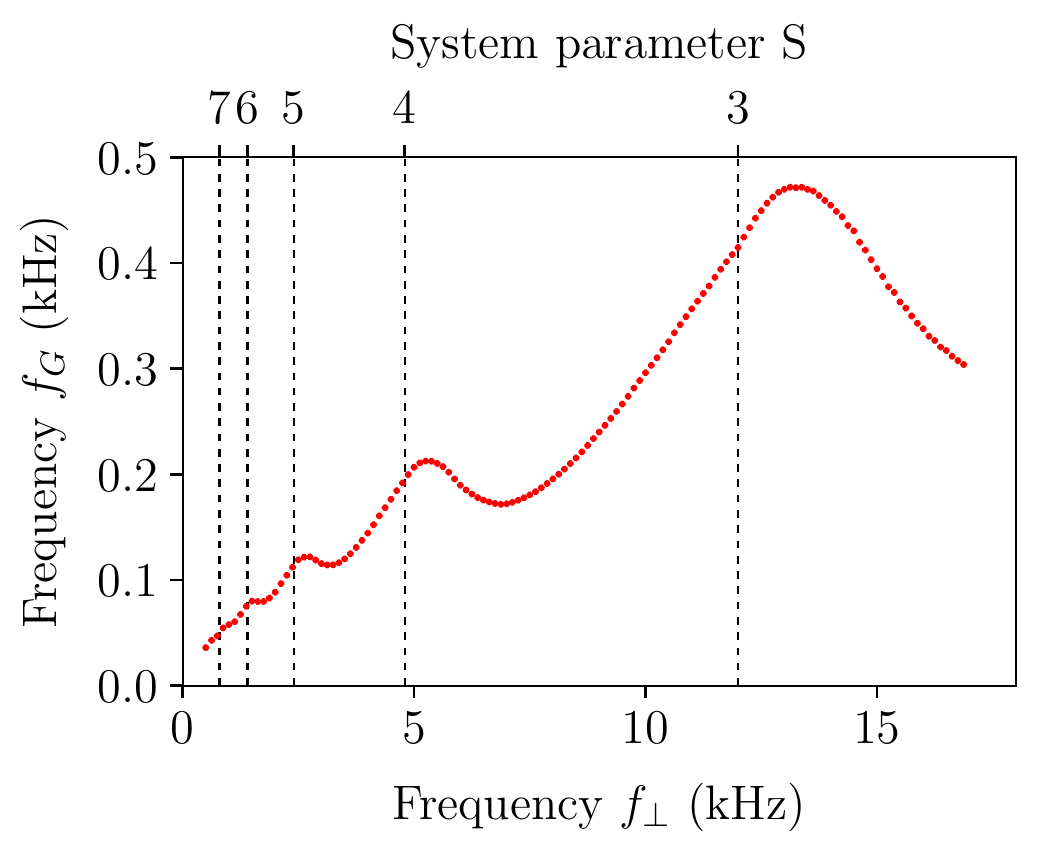}
	\caption{(color online) Frequency of the gapless Goldstone mode $f_G$ for fixed $f_\parallel = 96\,$Hz, $N_P = 1000$ and $1/(k_Fa) = -0.8 \rightarrow -0.9$; upper label: system paramter $S=\mu/\hbar \omega_\perp$.}
	\label{fig:resonanz}
\end{figure}

Figure \ref{fig:resonanz} shows the frequency of the gapless Goldstone mode for the above system over a wide range of $f_\perp$. One can see, that --on top of a global increasing trend\footnote{This global increase is due to the circumstance that the density of the condensate and thus the gap increases when $f_\perp$ is increased with fixed $f_\parallel$ and $N_P$.}-- $f_G$ exhibits a series of local maxima for different values of $f_\perp$. The distance of the maxima increases with increasing $f_\perp$ while at the same time the maxima become more pronounced. I.e., the maxima in $f_G$ occur less frequent but more pronounced when approaching higher values of $f_\perp$.

Furthermore, the dashed lines in Fig. \ref{fig:resonanz} indicate the positions of integer system parameter $S= \mu / \hbar \omega_\perp$ with $\omega_\perp = 2 \pi f_\perp$. These positions indicate trap parameters where the minimum of an atomic subband crosses the chemical potential (for a detailed description of the band structure see section \ref{ssec:single}). To be precise: The distance between two atomic subbands is $\hbar \omega_\perp$ and the minimum of the lowest subband is at $\varepsilon = \hbar \omega_\perp$. Therefore, the integer part of $S$ is the number of subbands that have a minimum below/at the chemical potential, i.e., the number of subbands crossing the chemical potential. One can see that every resonance closely follows such a point of integer system parameter $S$.

An explanation for this behavior can be given on the basis of the atypical BCS-BEC crossover \cite{Shanenko2012Atypical}: 
On the one hand, the atomic states closest to the chemical potential contribute strongest to the pairing. On the other hand, the states with the lowest quantum numbers $m_z$ which are located at the subband minima exhibit the strongest interaction matrix elements \cite{hannibal2015quench}. Therefore, each time an atomic subband crosses the chemical potential ($S=1,2,3,..$) the pairing is enhanced and the system is shifted towards the unitary point. Following section \ref{ssec:dynamics}, this results in a larger frequency of the gapless Goldstone mode.

However, the decrease of the impact of the resonances for increasing system parameter reflects the circumstance that for large $S$ several subbands contribute to the pairing while only a small fraction of the corresponding atomic states exhibits an enhanced coupling due to the resonance. Thus, for increasing $S$ the influence of the resonant states on the overall coupling decreases.

\subsection{Goldstone mode in the single-particle excitations}\label{ssec:single}

In this section we study the impact of the phase dynamics of the BCS gap on an experimentally more relevant physical quantity, the single-particle excitations of the condensate. An experimental investigation of the single-particle excitations has already been reported in \cite{Stewart2008Using} via RF-spectroscopy. Thus, they could provide a convenient access to the quench dynamics investigated here. Indeed, we will show that the gapless Goldstone mode is directly visible in the dynamics of the single-particle occupations and that it leads to a full inversion of the lowest-lying single-particle states. We will demonstrate this by investigating the effect of the phase dynamics on individual occupations as well as on the whole single-particle band structure. 

\begin{figure}[t]
	\centering
	\includegraphics[width=1\columnwidth]{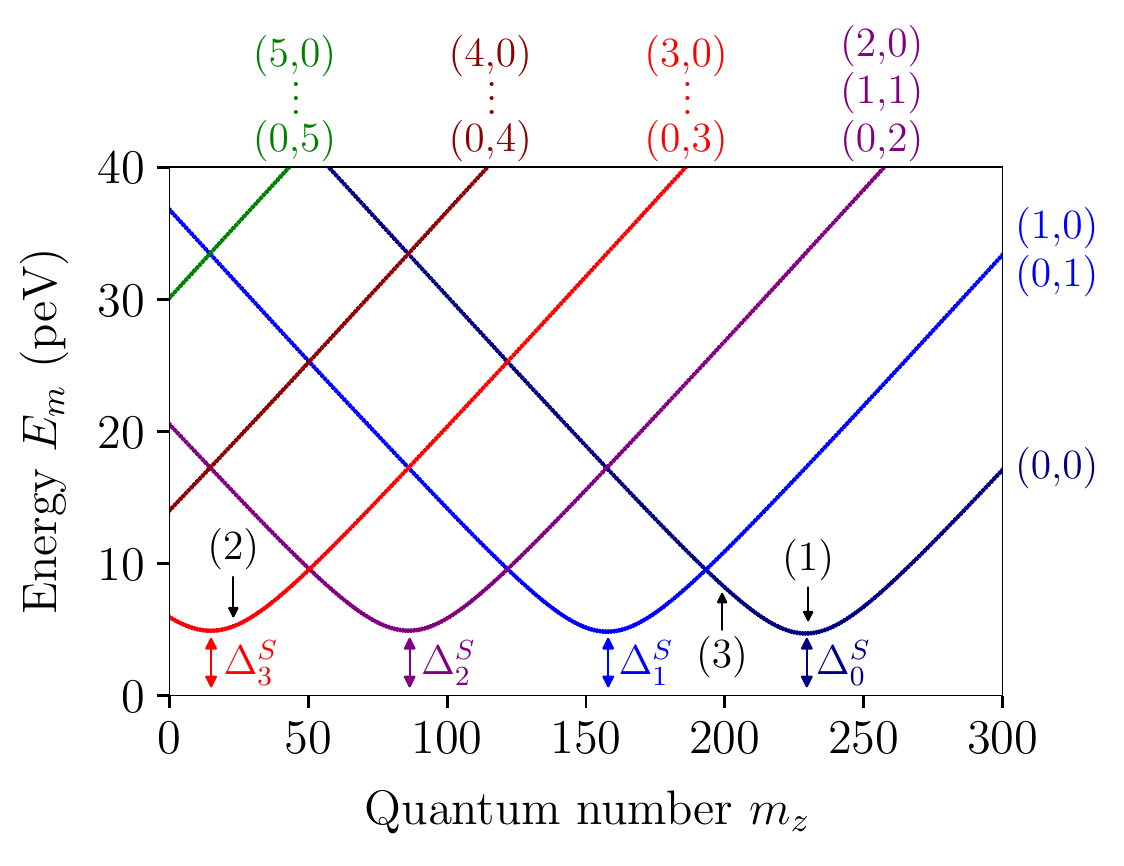}
	\caption{(color online) Single-particle energies for a strongly confined Fermi gas in a BCS phase with four atomic subbands crossing the chemical potential; $\Delta^{S}_i$ denotes the gap of the subbands with $m_x+m_y=i$ and $(m_x,m_y)$ denotes the subband index (see main text); the marked states (1), (2), (3) are discussed below; parameters: $f_\parallel = 56\,$Hz, $f_\perp = 4\,$kHz, $N=1700$ and $1/(k_Fa)=-0.9$.}
	\label{fig:Eq}
\end{figure}

In doing so, we focus on a cloud with the confinement frequencies given by $f_\parallel = 56\,$Hz and $f_\perp = 4\,$kHz, with $N_P=1700$ atoms in the trap and with an excitation of $1/(k_Fa)= -0.8 \rightarrow -0.9$. For such a cigar-shaped trap the atomic energies $\varepsilon_{m_x,m_y,m_z}:=\varepsilon_m$ are strongly separated with respect to $m_x$ and $m_y$ and comparatively dense with respect to $m_z$, i.e., they form subbands. The single-particle energies of Eq. \eqref{eq:Eq} inherit this band structure which can be seen in Fig. \ref{fig:Eq}.

There, a plot of the single-particle energies against the quantum number $m_z$ is shown for the system introduced above. One clearly observes several subbands each of which corresponds to certain sets of quantum numbers $(m_x, m_y)$, where --due to the cylindrical symmetry of the system-- each subband is $2(m_x+m_y+1)$ fold degenerate (the factor $2$ results from the degeneracy of the two single-particle branches corresponding to the two spin configurations)\footnote{Actually, subbands with different and not just interchanged quantum numbers $(m_x,m_y)$ are not exactly degenerate due to slightly different subband gaps $\Delta^{S}_m$ (on the order of $0.1\,$peV for the investigated systems). For example: The subbands $(0,3)$ and $(3,0)$ are exactly degenerate, whereas the subbands $(0,3)$ and $(1,2)$ are split by $\sim 0.5 \,$peV. However, in the presented plots and with the assumed experimental accuracy this splitting is not resolved.}. Furthermore, the four subbands with the lowest sets of quantum numbers $(m_x,m_y)$ show minima when the corresponding atomic subbands cross the chemical potential, i.e., at different values for $m_z$. The states located at the minima thus lie in close vicinity to the chemical potential and contribute strongly to the BCS pairing (the expectation values $\Delta^{\text{GS}}_{mm}$ corresponding to these states, i.e., the subband gaps, will be denoted as $\Delta^{S}_i$ with $i=m_x + m_y$ being the subband index; see Fig. \ref{fig:Eq}). The higher atomic subbands with $m_x + m_y \geq 4$ do not cross the chemical potential. The corresponding single-particle subbands therefore do not exhibit any minima.

Since the single-particle operators corresponding to the energies of Fig. \ref{fig:Eq} --i.e., those of Eqs. \eqref{eq:B-Transformation1} and \eqref{eq:B-Transformation2}-- are defined in the excitation picture all energy states of Fig. \ref{fig:Eq} are not occupied in the ground state before the quench. But, during the temporal evolution following the quench occupations of the order of 1 are created. We will show this explicitly for three particular single-particle states [marked as (1), (2) and (3) in Fig. \ref{fig:Eq}], one close to the minimum of the subband (1,2) ($E_m = 5.2\,$peV), one at the minimum of the subband (0,0) ($E_m = 4.7\,$peV) and one at a higher energy in the subband (0,0) ($E_m = 8.4\,$peV). Furthermore, we will identify the gapless Goldstone mode of the BCS gap in the corresponding dynamics.

\begin{figure}[t]
		\includegraphics[width = 1 \columnwidth]{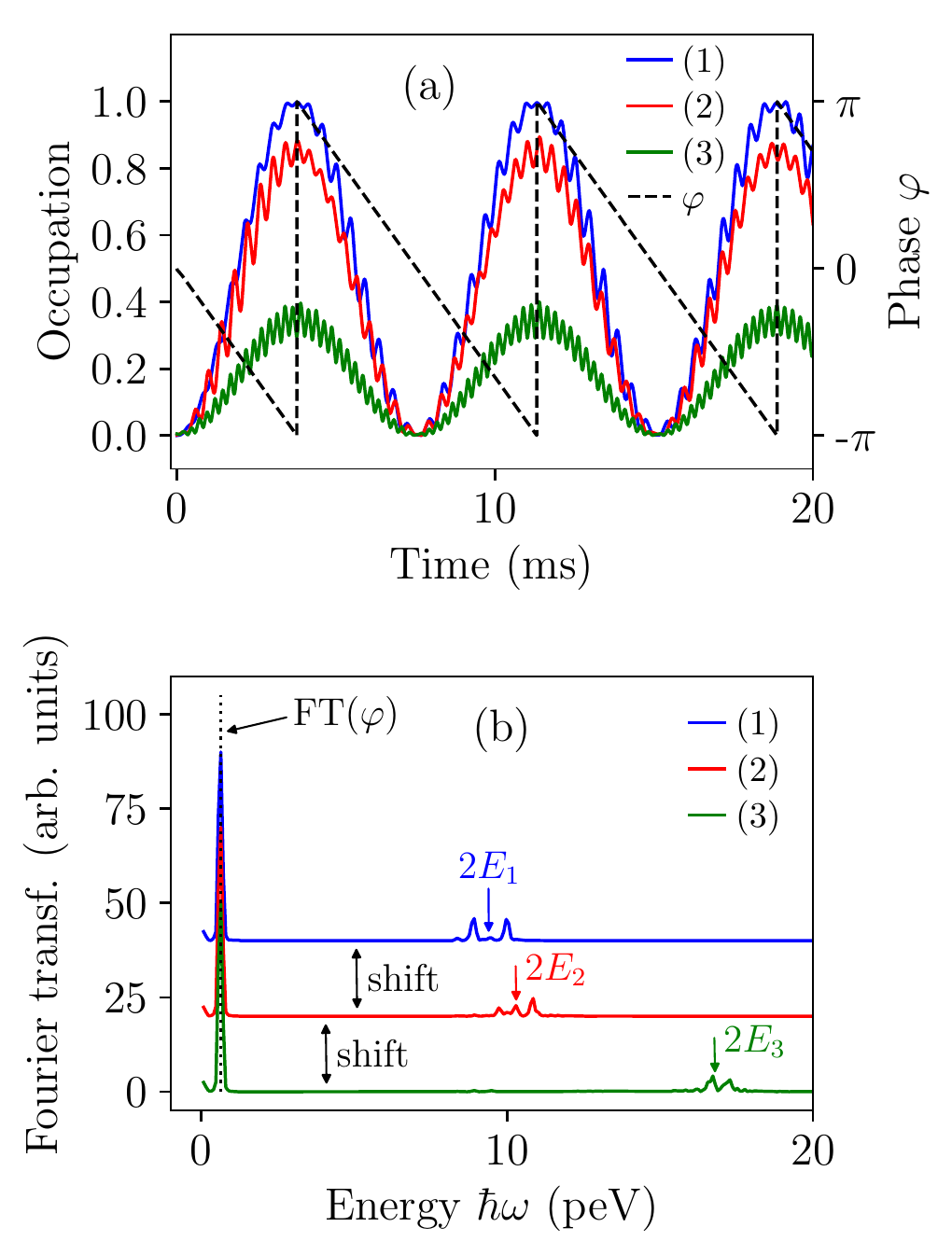}
		\caption{(color online) (a) Dynamics of three particular single-particle occupations, one state of higher energy (3) and two from subband minima (1), (2) (see Fig. \ref{fig:Eq}). (b) Fourier transform of the functions in (a).}
		\label{fig:gamma_t}
\end{figure}

The dynamics of the three single-particle occupations is shown in Fig. \ref{fig:gamma_t} (a) for the first 20$\,$ms after the quench. We clearly observe that all occupations oscillate in phase with one dominant low frequency. The states (1) (blue line) and (2) (red line) have a large amplitude of the order of 1 while the amplitude of state (3) (green line) is much smaller. Furthermore, the three occupations each exhibit an individual weak higher-frequency component which has the largest frequency for the state (3) of high energy. However, we find that the amplitude of the higher-frequency component increases with decreasing the scattering length, i.e., when entering the BCS regime with $1/(k_Fa) < -1$.

A comparison of the single-particle dynamics with the dynamics of the phase of the gap [Fig. \ref{fig:gamma_t} (a); dashed line] shows that the dominant low oscillation frequency originates from the Goldstone mode of the gap: The phase of the gap shows the same linear dynamics as in section \ref{ssec:dynamics} with a rate corresponding to the low-frequency part of the single-particle occupations. Thus, the gapless Goldstone mode is directly visible in the excitation dynamics of the condensate.

To investigate the single-particle dynamics in closer detail, Figure \ref{fig:gamma_t} (b) shows the Fourier spectrum of the data of Fig. \ref{fig:gamma_t} (a). Again, we observe that the dominant low frequencies of the occupations and the phase of the BCS gap exactly match. But, the origin of the higher frequencies in the excitation dynamics can now be seen as well: Besides the dominant low-frequency components each spectrum exhibits a series of weak peaks at approximately twice the energy of the corresponding single-particle state. I.e., the higher-frequency components result from an eigenoscillation of the single-particle occupations. In Ref. \cite{hannibal2015quench} a sum of all eigenoscillations was shown to result in the Higgs mode of the gap. The higher-frequency components can thus be understood as fragments of the Higgs mode of the BCS gap.

\begin{figure}[t]
		\includegraphics[width = 1\columnwidth]{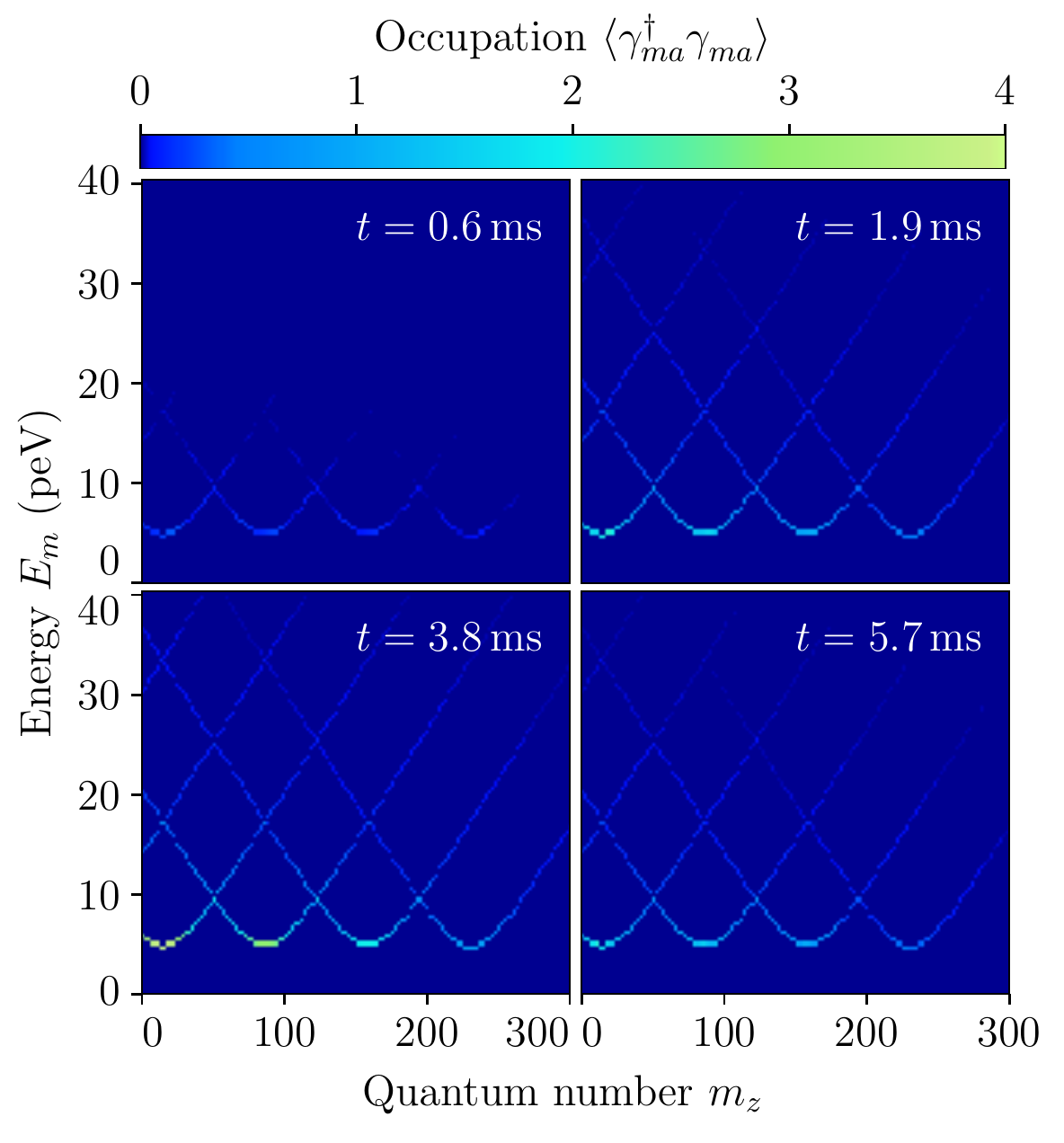}
		\caption{(color online) Single-particle occupations at different times after the interaction quench; parameters: See Fig. \ref{fig:Eq}. }
		\label{fig:spec_t}
\end{figure}

At last, --to analyze the impact of the gapless Goldstone mode on the whole single-particle spectrum-- Figure \ref{fig:spec_t} shows snapshots of the single-particle occupations plotted against the quantum number $m_z$ and the excitation energy $E_m$, like they could be measured by angle- and momentum-resolved RF spectroscopy \cite{Stewart2008Using}, for a series of time steps after the quench. The first snapshot corresponds to the time $t=0.6\,$ms and thus directly follows the quench. Here, the excitations are rather weak and can hardly be seen. However, going on in time we observe that all occupations increase in phase until the time $t=3.8\,$ms, where the maximum occupation of all states is reached. Afterwards the occupations decrease until the initial situation is reached again. Thus, Fig. \ref{fig:spec_t} is an illustration of the in-phase oscillation of all single-particle occupations due to the gapless Goldstone mode. 

In addition, --neglecting the contributions from the eigenoscillation-- the snapshot for $t=3.8\,$ms provides a map of the amplitude of the single-particle oscillations since here the dominant low-frequency part of all occupations exhibits its maximum value. On the basis of this amplitude map one observes, that the amplitude distribution shows a resonance behaviour: The amplitude is largest for states with low quantum number $m_z$ and low energy $E_m$ and decreases with increasing values of $m_z$ and $E_m$. In fact, directly at the minimum of the subbands with $m_x+m_y=3$ the oscillation amplitude is 4, decreasing by 1 for every next lower subband $m_x+m_y$. However, this dependence is due to the $(m_x+m_y+1)$-fold degeneracy of the subbands (only the particle-like excitations of the single-particle branch $a$ are shown). I.e., the oscillation amplitude of each individual single-particle occupation at a subband minimum is 1. Thus, the single-particle occupations at the subband minima exhibit a full inversion.

Concluding this section we can thus state: We have shown that the gapless Goldstone mode of an interaction-quenched ultracold Fermi gas directly couples to the single-particle occupations and leads to a full inversion of the lowest-lying states. An experimental access to the dynamical single-particle occupations would thus allow for a direct observation of the massless Goldstone Boson predicted by the Goldstone theorem. However, we want to remark that an application of RF-spectroscopy --a state-of-the-art experimental access to the single-particle excitations \cite{Stewart2008Using,Stewart2010Verification}-- to the dynamical situation is restricted to the observation of the inhomogeneous phase dynamics. It turns out that it does not contain any signature of the gapless homogeneous Goldstone mode (see appendix). Therefore, at least a modification of this experimental technique would be required to observe the gapless Goldstone mode via the single-particle excitations.

\section{Conclusion}\label{sec:conclusions}

In conlusion, we have calculated the dynamics of a confined ultracold $^6$Li gas at $T=0$ induced by an interaction quench on the BCS side of the BCS-BEC crossover. We used a full dynamical BdG approach to set up and solve the equations of motion for the single-particle occupations and coherences. In doing so, we have shown that the interaction quench excites a low-energy linear dynamics of the complex phase of the BCS gap, i.e., a Goldstone mode. We have analyzed this Goldstone mode over a wide range of parameters showing that its excitation spectrum is gapless and that its main frequency is not fixed by the trap frequencies but that it is determined by the details of the quench. Furthermore, we found that the atypical BCS-BEC crossover leads to resonances in the gapless Goldstone mode. Finally, we investigated the impact of the gapless Goldstone mode on the single-particle occupations. We have shown that it leads to an in-phase oscillation of the whole single-particle spectrum with a full inversion of the lowest-lying single-particle states which could provide an experimental access to the gapless homogeneous Goldstone mode.

\appendix*

\section{Gapless Goldstone mode and RF spectroscopy}\label{sec:appendix}

One way to study the single-particle occupations in experiment is RF spectroscopy as was shown in Ref. \cite{Stewart2008Using}. There, a first direct measurement of the single-particle excitations via RF spectroscopy was achieved for a thermal superfluid gas of ultracold $^{40}$K in the BCS-BEC crossover. An application of RF spectroscopy to the dynamical situation could thus allow for a direct observation of the Goldstone mode without coupling it to the trap. This could be achieved, e.g., via a pump-probe like experimental setup: By introducing a delay time $t$ between the quench (the \enquote{pump pulse}) and the actual RF measurement (the \enquote{probe pulse}) a time-resolved single-particle spectrum like in Fig. \ref{fig:spec_t} could be obtained.  However, in the following we will show that RF signals give a direct measurement of the single-particle occupations only if no single-particle coherences are present. In contrast, we will demonstrate that --for our case of a coherent evolution of the condensate-- the single-particle coherences cancel the signature of the gapless Goldstone mode in the RF signal and thus prohibit its direct observation via RF spectroscopy.
\medskip

The basic principle of RF spectroscopy applied to ultracold Fermi gases is to optically excite the atoms from one of the two hyperfine states of the condensate, i.e., the state denoted as $|k\uparrow\rangle$, to a third hyperfine state which is not involved in the BCS pairing (following Ref. \cite{Ketterle2008Making} we will denote this state as $|k\sigma\rangle$ with $\sigma=3$; depending on the atom species used the actual spin of the corresponding hyperfine state will be different though). Then, the resulting occupations of the third hyperfine state can be used --at least in the absence of coherences between the single-particle states-- as a direct measure of the corresponding single-particle occupations (see below and \cite{Ketterle2008Making}). 

As stated above, an RF excitation results in a simultaneous spin flip of all atoms in the cloud which is described by the operator \cite{Ketterle2008Making}
\begin{equation} \label{eq:V_RF}
\hat{V} = V_0 \sum\limits_k \left(c^{\dagger}_{k3}c_{k\uparrow} + c^{\dagger}_{k\uparrow} c_{k3}\right),
\end{equation}
with $c^{\dagger}_{k\sigma}$ ($c^{ }_{k\sigma}$)  creating (annihilating) one atom in the state with quantum number $k$ and spin index $\sigma \in \{\uparrow,\downarrow,3\}$. Here, we assume the excitation to be orthogonal with respect to the quantum number $k$, i.e., we only consider transitions $|k\uparrow\rangle \rightarrow |k^\prime 3\rangle$ with $k=k^\prime$. Strictly speaking this applies only to very large systems where $k$ corresponds to the wave number and if we furthermore assume $\hbar k_{RF} \ll \hbar k$, i.e., that the momentum of the photons is much smaller than the momentum of the atoms. However, in Anderson Approximation transitions with $k\neq k^\prime$ do not contribute to the RF signal since all nondiagonal single-particle expectation values vanish (see below). Thus, we can apply Eq. \eqref{eq:V_RF} to our current situation.

Following \cite{Ketterle2008Making}, we use
\begin{eqnarray}\label{eq:c_Bog}
c_{k\uparrow} = u_k \gamma_{ka}^\dagger - v_k \gamma_{kb}
\end{eqnarray}
and we furthermore assume the third hyperfine state to be initially empty, i.e., $c_{k3}|\Psi(t) \rangle = 0$ with $|\Psi(t) \rangle$ the state of the condensate before the RF pulse. This yields:
\begin{eqnarray}\label{eq:V_RF_Bog}
\hat{V} = V_0 \sum\limits_k c^{\dagger}_{k3}\left(u_k \gamma_{ka}^\dagger - v_k \gamma_{kb}\right).
\end{eqnarray}
Therefore, the RF excitation transfers single atoms to the state $|k3\rangle$ by creating a quasiparticle in the single-particle state $ka$ and destroying one in state $kb$. However, we are interested in the occupations of state $|k3\rangle$ after the excitation, i.e., we have to investigate the transitions governed by the matrix elements
\begin{eqnarray}\label{eq:transition}
M_k = \langle f|V|\Psi(t)\rangle,
\end{eqnarray}
where $|\Psi(t)\rangle = | \gamma(t) \rangle |0\rangle_{3}$ is composed of the quasiparticle contribution  $|\gamma(t)\rangle$  and the vacuum of the third hyperfine state $|0\rangle_{3}$. Since the exact quasiparticle configuration after the RF pulse is not relevant, the final state of the transition $|f\rangle$ needs to take into account all possible end states for the quasiparticles, i.e., 
\begin{equation} \label{eq:final}
|f\rangle= \sum\limits_{k^\prime} \left(u_{k^\prime} \gamma_{k^\prime a}^\dagger - v_{k^\prime} \gamma_{k^\prime b}\right) |\gamma(t)\rangle |k3\rangle.
\end{equation}
Inserting this into Eq. \eqref{eq:transition} and keeping in mind that in Anderson approximation $\langle \gamma_{ma/b}^\dagger \gamma_{na/b} \rangle = \langle \gamma_{ma}^\dagger \gamma_{nb}^\dagger \rangle = 0$ for $m\neq n$ and that $\langle \gamma_{ma}^\dagger \gamma_{ma} \rangle = \langle \gamma_{mb}^\dagger \gamma_{mb} \rangle$ \cite{hannibal2015quench} we obtain
\begin{equation} \label{eq:transition_final}
M_k \sim  (u_{k}^2 - v_{k}^2) \langle \gamma_{ka}^\dagger \gamma_{ka} \rangle - 2u_k v_k\text{Re}\left( \langle \gamma_{ka}^\dagger \gamma_{kb}^\dagger \rangle \right)+ v_{k}^2.
\end{equation}
Therefore, the population in the state $|k3\rangle$ created by the RF pulse is a direct measure of the single-particle occupations if the single-particle coherences $\langle \gamma_{ka}^\dagger \gamma_{kb}^\dagger \rangle$ vanish. This is the case for every thermal state of the condensate. I.e., --in that case-- a momentum- and energy-resolved measurement of the occupation of the third hyperfine state maps the single-particle band structure (cf. \cite{Stewart2008Using}).

However, if single-particle coherences are present, they may interfere with the signal from the occupations and prohibit an observation of the latter. This is the case for our situation as can be seen by calculating $M_k$ directly from Eq. \eqref{eq:V_RF} The quasiparticle part of the transition matrix elements $M_k$ is basically given by the occupation of the atomic state $|k\uparrow\rangle$, i.e.,
\begin{equation}
M_k \sim \langle c^\dagger_{k\uparrow} c_{k\uparrow} \rangle
\end{equation}
Therefore, the overall transition matrix element considering all transtitions to the third hyperfine state yields
\begin{equation}\label{eq:transition_total}
M_{\text{total}} \sim \sum\limits_k \langle c^\dagger_{k\uparrow} c_{k\uparrow} \rangle = N_{P\uparrow}.
\end{equation}
However, $N_{P\uparrow}$ is a conserved quantity during the free quench dynamics, i.e., during the dynamics before the RF pulse. Therefore, $M_{\text{total}}$ and thus the overall RF-induced occupations of the third hyperfine state do not depend on the actual time of the RF measurement. This implies that the contribution from the Goldstone mode in the occupations has to be canceled out by a corresponding contribution in the single-particle coherences since all single-paricle occupations oscillate in phase with respect to the gapless Goldstone mode. Therefore, every individual $M_k$ is a constant with respect to the Goldstone mode. In that sense, particle conservation prohibits an observation of the gapless Goldstone mode via RF spectroscopy.

Indeed, our numerical data confirm that the contribution of the single-particle coherences to Eq. \eqref{eq:transition_final} exactly cancels out the signal of the Goldstone mode from the single-particle occupations. This confirms that the Goldstone mode in the present form is not visible via RF spectroscopy.

However, the same applies to the real-space dynamics of the condensate: For the same reasons as stated above the gapless homogeneous Goldstone mode is not visible in the atomic density $\rho(\vec{r},t)$ of the cloud. With Eqs. \eqref{eq:B-Transformation1}-\eqref{eq:B-Transformation2} one directly obtains
\begin{align} \label{eq:rho}
\rho(\vec{r},t) &= \sum\limits_\sigma \langle \hat{\Psi}^\dagger_\sigma(\vec{r},t) \hat{\Psi}_\sigma(\vec{r},t) \rangle \nonumber \\
&= 2 \sum \limits_k \big[ (u_{k}^2 - v_{k}^2) \langle \gamma_{ka}^\dagger \gamma_{ka} \rangle \nonumber \\ 
& \hspace{1cm} - 2u_k v_k\text{Re}\left(\langle \gamma_{ka}^\dagger \gamma_{kb}^\dagger \rangle \right)+ v_{k}^2\big] |\varphi_k(\vec{r})|^2,
\end{align}
which has the same structure as Eq. \eqref{eq:transition_final}. Therefore, the gapless homogeneous Goldstone mode does not couple to the real-space dynamics of the cloud as already stated in section \ref{ssec:dynamics}. However, the resemblence of Eqs. \eqref{eq:rho} and \eqref{eq:transition_final} also implies that the Goldstone mode is visible in the RF signal for the case of inhomogeneous excitations. There, the symmetry between the single-particle excitations and coherences preventing the gapless homogeneous Goldstone mode from observation must be broken to allow for a collective oscillation of the cloud. Therefore, we state that in our case of a homogeneous excitation the Goldstone mode is visible neither in the single-particle excitations nor in the real-space dynamics of the cloud. But, for the same reason it must be visible in both quantities for the case of inhomogeneous excitations.
 



\begin{thebibliography}{44}%
\makeatletter
\providecommand \@ifxundefined [1]{%
 \@ifx{#1\undefined}
}%
\providecommand \@ifnum [1]{%
 \ifnum #1\expandafter \@firstoftwo
 \else \expandafter \@secondoftwo
 \fi
}%
\providecommand \@ifx [1]{%
 \ifx #1\expandafter \@firstoftwo
 \else \expandafter \@secondoftwo
 \fi
}%
\providecommand \natexlab [1]{#1}%
\providecommand \enquote  [1]{``#1''}%
\providecommand \bibnamefont  [1]{#1}%
\providecommand \bibfnamefont [1]{#1}%
\providecommand \citenamefont [1]{#1}%
\providecommand \href@noop [0]{\@secondoftwo}%
\providecommand \href [0]{\begingroup \@sanitize@url \@href}%
\providecommand \@href[1]{\@@startlink{#1}\@@href}%
\providecommand \@@href[1]{\endgroup#1\@@endlink}%
\providecommand \@sanitize@url [0]{\catcode `\\12\catcode `\$12\catcode
  `\&12\catcode `\#12\catcode `\^12\catcode `\_12\catcode `\%12\relax}%
\providecommand \@@startlink[1]{}%
\providecommand \@@endlink[0]{}%
\providecommand \url  [0]{\begingroup\@sanitize@url \@url }%
\providecommand \@url [1]{\endgroup\@href {#1}{\urlprefix }}%
\providecommand \urlprefix  [0]{URL }%
\providecommand \Eprint [0]{\href }%
\providecommand \doibase [0]{http://dx.doi.org/}%
\providecommand \selectlanguage [0]{\@gobble}%
\providecommand \bibinfo  [0]{\@secondoftwo}%
\providecommand \bibfield  [0]{\@secondoftwo}%
\providecommand \translation [1]{[#1]}%
\providecommand \BibitemOpen [0]{}%
\providecommand \bibitemStop [0]{}%
\providecommand \bibitemNoStop [0]{.\EOS\space}%
\providecommand \EOS [0]{\spacefactor3000\relax}%
\providecommand \BibitemShut  [1]{\csname bibitem#1\endcsname}%
\let\auto@bib@innerbib\@empty
\bibitem [{\citenamefont {Giorgini}\ \emph {et~al.}(2008)\citenamefont
  {Giorgini}, \citenamefont {Pitaevskii},\ and\ \citenamefont
  {Stringari}}]{Giorgini2008Theory}%
  \BibitemOpen
  \bibfield  {author} {\bibinfo {author} {\bibfnamefont {S.}~\bibnamefont
  {Giorgini}}, \bibinfo {author} {\bibfnamefont {L.}~\bibnamefont
  {Pitaevskii}}, \ and\ \bibinfo {author} {\bibfnamefont {S.}~\bibnamefont
  {Stringari}},\ }\bibfield  {title} {\enquote {\bibinfo {title} {Theory of
  ultracold atomic {F}ermi gases},}\ }\href@noop {} {\bibfield  {journal}
  {\bibinfo  {journal} {Rev. Mod. Phys.}\ }\textbf {\bibinfo {volume} {80}},\
  \bibinfo {pages} {1215--1274} (\bibinfo {year} {2008})}\BibitemShut {NoStop}%
\bibitem [{\citenamefont {Bloch}\ \emph {et~al.}(2008)\citenamefont {Bloch},
  \citenamefont {Dalibard},\ and\ \citenamefont {Zwerger}}]{Bloch2008Many}%
  \BibitemOpen
  \bibfield  {author} {\bibinfo {author} {\bibfnamefont {I.}~\bibnamefont
  {Bloch}}, \bibinfo {author} {\bibfnamefont {J.}~\bibnamefont {Dalibard}}, \
  and\ \bibinfo {author} {\bibfnamefont {W.}~\bibnamefont {Zwerger}},\
  }\bibfield  {title} {\enquote {\bibinfo {title} {Many-body physics with
  ultracold gases},}\ }\href@noop {} {\bibfield  {journal} {\bibinfo  {journal}
  {Rev. Mod. Phys.}\ }\textbf {\bibinfo {volume} {80}},\ \bibinfo {pages} {885}
  (\bibinfo {year} {2008})}\BibitemShut {NoStop}%
\bibitem [{\citenamefont {Weinberg}(1996)}]{weinberg1996quantum}%
  \BibitemOpen
  \bibfield  {author} {\bibinfo {author} {\bibfnamefont {Steven}\ \bibnamefont
  {Weinberg}},\ }\href@noop {} {\emph {\bibinfo {title} {The quantum theory of
  fields}}},\ Vol.~\bibinfo {volume} {2}\ (\bibinfo  {publisher} {Cambridge
  university press},\ \bibinfo {year} {1996})\ pp.\ \bibinfo {pages}
  {332--352}\BibitemShut {NoStop}%
\bibitem [{\citenamefont {Higgs}(1964)}]{Higgs1964Broken}%
  \BibitemOpen
  \bibfield  {author} {\bibinfo {author} {\bibfnamefont {P.~W.}\ \bibnamefont
  {Higgs}},\ }\bibfield  {title} {\enquote {\bibinfo {title} {Broken symmetries
  and the masses of gauge bosons},}\ }\href@noop {} {\bibfield  {journal}
  {\bibinfo  {journal} {Phys. Rev. Lett.}\ }\textbf {\bibinfo {volume} {13}},\
  \bibinfo {pages} {508--509} (\bibinfo {year} {1964})}\BibitemShut {NoStop}%
\bibitem [{\citenamefont {Burgess}(2000)}]{burgess2000goldstone}%
  \BibitemOpen
  \bibfield  {author} {\bibinfo {author} {\bibfnamefont {C.~P.}\ \bibnamefont
  {Burgess}},\ }\bibfield  {title} {\enquote {\bibinfo {title} {Goldstone and
  pseudo-{G}oldstone bosons in nuclear, particle and condensed-matter
  physics},}\ }\href@noop {} {\bibfield  {journal} {\bibinfo  {journal} {Phys.
  Reports}\ }\textbf {\bibinfo {volume} {330}},\ \bibinfo {pages} {193--261}
  (\bibinfo {year} {2000})}\BibitemShut {NoStop}%
\bibitem [{\citenamefont {Paulson}\ \emph {et~al.}(1973)\citenamefont
  {Paulson}, \citenamefont {Johnson},\ and\ \citenamefont
  {Wheatley}}]{Paulson1973Propagation}%
  \BibitemOpen
  \bibfield  {author} {\bibinfo {author} {\bibfnamefont {D.}~\bibnamefont
  {Paulson}}, \bibinfo {author} {\bibfnamefont {R.}~\bibnamefont {Johnson}}, \
  and\ \bibinfo {author} {\bibfnamefont {J.}~\bibnamefont {Wheatley}},\
  }\bibfield  {title} {\enquote {\bibinfo {title} {Propagation of collisionless
  sound in normal and extraordinary phases of liquid $^3${H}e below
  3$\,$m{K}},}\ }\href@noop {} {\bibfield  {journal} {\bibinfo  {journal}
  {Phys. Rev. Lett.}\ }\textbf {\bibinfo {volume} {30}},\ \bibinfo {pages}
  {829} (\bibinfo {year} {1973})}\BibitemShut {NoStop}%
\bibitem [{\citenamefont {Lawson}\ \emph {et~al.}(1973)\citenamefont {Lawson},
  \citenamefont {Gully}, \citenamefont {Goldstein}, \citenamefont
  {Richardson},\ and\ \citenamefont {Lee}}]{Lawson1973Attenuation}%
  \BibitemOpen
  \bibfield  {author} {\bibinfo {author} {\bibfnamefont {D.}~\bibnamefont
  {Lawson}}, \bibinfo {author} {\bibfnamefont {W.}~\bibnamefont {Gully}},
  \bibinfo {author} {\bibfnamefont {S.}~\bibnamefont {Goldstein}}, \bibinfo
  {author} {\bibfnamefont {R.}~\bibnamefont {Richardson}}, \ and\ \bibinfo
  {author} {\bibfnamefont {D.}~\bibnamefont {Lee}},\ }\bibfield  {title}
  {\enquote {\bibinfo {title} {Attenuation of zero sound and the
  low-temperature transitions in liquid $^3${H}e},}\ }\href@noop {} {\bibfield
  {journal} {\bibinfo  {journal} {Phys. Rev. Lett.}\ }\textbf {\bibinfo
  {volume} {30}},\ \bibinfo {pages} {541} (\bibinfo {year} {1973})}\BibitemShut
  {NoStop}%
\bibitem [{\citenamefont {Anderson}(1958)}]{anderson1958random}%
  \BibitemOpen
  \bibfield  {author} {\bibinfo {author} {\bibfnamefont {P.~W.}\ \bibnamefont
  {Anderson}},\ }\bibfield  {title} {\enquote {\bibinfo {title} {Random-phase
  approximation in the theory of superconductivity},}\ }\href@noop {}
  {\bibfield  {journal} {\bibinfo  {journal} {Phys. Rev.}\ }\textbf {\bibinfo
  {volume} {112}},\ \bibinfo {pages} {1900} (\bibinfo {year}
  {1958})}\BibitemShut {NoStop}%
\bibitem [{\citenamefont {Pekker}\ and\ \citenamefont
  {Varma}(2015)}]{pekker2015amplitude}%
  \BibitemOpen
  \bibfield  {author} {\bibinfo {author} {\bibfnamefont {David}\ \bibnamefont
  {Pekker}}\ and\ \bibinfo {author} {\bibfnamefont {CM}~\bibnamefont {Varma}},\
  }\bibfield  {title} {\enquote {\bibinfo {title} {Amplitude/higgs modes in
  condensed matter physics},}\ }\href@noop {} {\bibfield  {journal} {\bibinfo
  {journal} {Annu. Rev. Condens. Matter Phys.}\ }\textbf {\bibinfo {volume}
  {6}},\ \bibinfo {pages} {269--297} (\bibinfo {year} {2015})}\BibitemShut
  {NoStop}%
\bibitem [{\citenamefont {Bissbort}\ \emph {et~al.}(2011)\citenamefont
  {Bissbort}, \citenamefont {G{\"o}tze}, \citenamefont {Li}, \citenamefont
  {Heinze}, \citenamefont {Krauser}, \citenamefont {Weinberg}, \citenamefont
  {Becker}, \citenamefont {Sengstock},\ and\ \citenamefont
  {Hofstetter}}]{bissbort2011detecting}%
  \BibitemOpen
  \bibfield  {author} {\bibinfo {author} {\bibfnamefont {U.}~\bibnamefont
  {Bissbort}}, \bibinfo {author} {\bibfnamefont {S.}~\bibnamefont {G{\"o}tze}},
  \bibinfo {author} {\bibfnamefont {Y.}~\bibnamefont {Li}}, \bibinfo {author}
  {\bibfnamefont {J.}~\bibnamefont {Heinze}}, \bibinfo {author} {\bibfnamefont
  {J.~S}\ \bibnamefont {Krauser}}, \bibinfo {author} {\bibfnamefont
  {M.}~\bibnamefont {Weinberg}}, \bibinfo {author} {\bibfnamefont
  {C.}~\bibnamefont {Becker}}, \bibinfo {author} {\bibfnamefont
  {K.}~\bibnamefont {Sengstock}}, \ and\ \bibinfo {author} {\bibfnamefont
  {W.}~\bibnamefont {Hofstetter}},\ }\bibfield  {title} {\enquote {\bibinfo
  {title} {Detecting the amplitude mode of strongly interacting lattice bosons
  by {B}ragg scattering},}\ }\href@noop {} {\bibfield  {journal} {\bibinfo
  {journal} {Phys. Rev. Lett.}\ }\textbf {\bibinfo {volume} {106}},\ \bibinfo
  {pages} {205303} (\bibinfo {year} {2011})}\BibitemShut {NoStop}%
\bibitem [{\citenamefont {Endres}\ \emph {et~al.}(2012)\citenamefont {Endres},
  \citenamefont {Fukuhara}, \citenamefont {Pekker}, \citenamefont {Cheneau},
  \citenamefont {Schau{\ss}}, \citenamefont {Gross}, \citenamefont {Demler},
  \citenamefont {Kuhr},\ and\ \citenamefont
  {Bloch}}]{Endres2012The/Higgs/amplitude}%
  \BibitemOpen
  \bibfield  {author} {\bibinfo {author} {\bibfnamefont {M.}~\bibnamefont
  {Endres}}, \bibinfo {author} {\bibfnamefont {T.}~\bibnamefont {Fukuhara}},
  \bibinfo {author} {\bibfnamefont {D.}~\bibnamefont {Pekker}}, \bibinfo
  {author} {\bibfnamefont {M.}~\bibnamefont {Cheneau}}, \bibinfo {author}
  {\bibfnamefont {P.}~\bibnamefont {Schau{\ss}}}, \bibinfo {author}
  {\bibfnamefont {C.}~\bibnamefont {Gross}}, \bibinfo {author} {\bibfnamefont
  {E.}~\bibnamefont {Demler}}, \bibinfo {author} {\bibfnamefont
  {S.}~\bibnamefont {Kuhr}}, \ and\ \bibinfo {author} {\bibfnamefont
  {I.}~\bibnamefont {Bloch}},\ }\bibfield  {title} {\enquote {\bibinfo {title}
  {The '{H}iggs' amplitude mode at the two-dimensional superfluid/{M}ott
  insulator transition},}\ }\href@noop {} {\bibfield  {journal} {\bibinfo
  {journal} {Nature}\ }\textbf {\bibinfo {volume} {487}},\ \bibinfo {pages}
  {454--458} (\bibinfo {year} {2012})}\BibitemShut {NoStop}%
\bibitem [{\citenamefont {Matsunaga}\ \emph {et~al.}(2013)\citenamefont
  {Matsunaga}, \citenamefont {Hamada}, \citenamefont {Makise}, \citenamefont
  {Uzawa}, \citenamefont {Terai}, \citenamefont {Wang},\ and\ \citenamefont
  {Shimano}}]{Matsunaga2013Higgs}%
  \BibitemOpen
  \bibfield  {author} {\bibinfo {author} {\bibfnamefont {R.}~\bibnamefont
  {Matsunaga}}, \bibinfo {author} {\bibfnamefont {Y.~I.}\ \bibnamefont
  {Hamada}}, \bibinfo {author} {\bibfnamefont {K.}~\bibnamefont {Makise}},
  \bibinfo {author} {\bibfnamefont {Y.}~\bibnamefont {Uzawa}}, \bibinfo
  {author} {\bibfnamefont {H.}~\bibnamefont {Terai}}, \bibinfo {author}
  {\bibfnamefont {Z.}~\bibnamefont {Wang}}, \ and\ \bibinfo {author}
  {\bibfnamefont {R.}~\bibnamefont {Shimano}},\ }\bibfield  {title} {\enquote
  {\bibinfo {title} {{H}iggs amplitude mode in the {BCS} superconductors {N}b
  1-x {T}i x n induced by terahertz pulse excitation},}\ }\href@noop {}
  {\bibfield  {journal} {\bibinfo  {journal} {Phys. Rev. Lett.}\ }\textbf
  {\bibinfo {volume} {111}},\ \bibinfo {pages} {057002} (\bibinfo {year}
  {2013})}\BibitemShut {NoStop}%
\bibitem [{\citenamefont {Matsunaga}\ \emph {et~al.}(2014)\citenamefont
  {Matsunaga}, \citenamefont {Tsuji}, \citenamefont {Fujita}, \citenamefont
  {Sugioka}, \citenamefont {Makise}, \citenamefont {Uzawa}, \citenamefont
  {Terai}, \citenamefont {Wang}, \citenamefont {Aoki},\ and\ \citenamefont
  {Shimano}}]{Matsunaga2014Light}%
  \BibitemOpen
  \bibfield  {author} {\bibinfo {author} {\bibfnamefont {R.}~\bibnamefont
  {Matsunaga}}, \bibinfo {author} {\bibfnamefont {N.}~\bibnamefont {Tsuji}},
  \bibinfo {author} {\bibfnamefont {H.}~\bibnamefont {Fujita}}, \bibinfo
  {author} {\bibfnamefont {A.}~\bibnamefont {Sugioka}}, \bibinfo {author}
  {\bibfnamefont {K.}~\bibnamefont {Makise}}, \bibinfo {author} {\bibfnamefont
  {Y.}~\bibnamefont {Uzawa}}, \bibinfo {author} {\bibfnamefont
  {H.}~\bibnamefont {Terai}}, \bibinfo {author} {\bibfnamefont
  {Z.}~\bibnamefont {Wang}}, \bibinfo {author} {\bibfnamefont {H.}~\bibnamefont
  {Aoki}}, \ and\ \bibinfo {author} {\bibfnamefont {R.}~\bibnamefont
  {Shimano}},\ }\bibfield  {title} {\enquote {\bibinfo {title} {Light-induced
  collective pseudospin precession resonating with {H}iggs mode in a
  superconductor},}\ }\href@noop {} {\bibfield  {journal} {\bibinfo  {journal}
  {Science}\ }\textbf {\bibinfo {volume} {345}},\ \bibinfo {pages} {1145--1149}
  (\bibinfo {year} {2014})}\BibitemShut {NoStop}%
\bibitem [{\citenamefont {Barankov}\ \emph {et~al.}(2004)\citenamefont
  {Barankov}, \citenamefont {Levitov},\ and\ \citenamefont
  {Spivak}}]{Barankov2004Collective}%
  \BibitemOpen
  \bibfield  {author} {\bibinfo {author} {\bibfnamefont {R.A.}~\bibnamefont
  {Barankov}}, \bibinfo {author} {\bibfnamefont {L.S.}~\bibnamefont {Levitov}}, \
  and\ \bibinfo {author} {\bibfnamefont {B.Z.}~\bibnamefont {Spivak}},\
  }\bibfield  {title} {\enquote {\bibinfo {title} {Collective {R}abi
  oscillations and solitons in a time-dependent {BCS} pairing problem},}\
  }\href@noop {} {\bibfield  {journal} {\bibinfo  {journal} {Phys. Rev. Lett.}\
  }\textbf {\bibinfo {volume} {93}},\ \bibinfo {pages} {160401} (\bibinfo
  {year} {2004})}\BibitemShut {NoStop}%
\bibitem [{\citenamefont {Barankov}\ and\ \citenamefont
  {Levitov}(2006)}]{Barankov2006Synchronization}%
  \BibitemOpen
  \bibfield  {author} {\bibinfo {author} {\bibfnamefont {R.~A.}\ \bibnamefont
  {Barankov}}\ and\ \bibinfo {author} {\bibfnamefont {L.~S.}\ \bibnamefont
  {Levitov}},\ }\bibfield  {title} {\enquote {\bibinfo {title} {Synchronization
  in the {BCS} pairing dynamics as a critical phenomenon},}\ }\href@noop {}
  {\bibfield  {journal} {\bibinfo  {journal} {Phys. Rev. Lett.}\ }\textbf
  {\bibinfo {volume} {96}},\ \bibinfo {pages} {230403} (\bibinfo {year}
  {2006})}\BibitemShut {NoStop}%
\bibitem [{\citenamefont {Yuzbashyan}\ \emph {et~al.}(2006)\citenamefont
  {Yuzbashyan}, \citenamefont {Tsyplyatyev},\ and\ \citenamefont
  {Altshuler}}]{Yuzbashyan2006Relaxation}%
  \BibitemOpen
  \bibfield  {author} {\bibinfo {author} {\bibfnamefont {E.~A.}\ \bibnamefont
  {Yuzbashyan}}, \bibinfo {author} {\bibfnamefont {O.}~\bibnamefont
  {Tsyplyatyev}}, \ and\ \bibinfo {author} {\bibfnamefont {B.~L.}\ \bibnamefont
  {Altshuler}},\ }\bibfield  {title} {\enquote {\bibinfo {title} {Relaxation
  and persistent oscillations of the order parameter in fermionic
  condensates},}\ }\href@noop {} {\bibfield  {journal} {\bibinfo  {journal}
  {Phys. Rev. Lett.}\ }\textbf {\bibinfo {volume} {96}},\ \bibinfo {pages}
  {097005} (\bibinfo {year} {2006})}\BibitemShut {NoStop}%
\bibitem [{\citenamefont {Dzero}\ \emph {et~al.}(2007)\citenamefont {Dzero},
  \citenamefont {Yuzbashyan}, \citenamefont {Altshuler},\ and\ \citenamefont
  {Coleman}}]{Dzero2007Spectroscopic}%
  \BibitemOpen
  \bibfield  {author} {\bibinfo {author} {\bibfnamefont {M.}~\bibnamefont
  {Dzero}}, \bibinfo {author} {\bibfnamefont {E.A.}~\bibnamefont {Yuzbashyan}},
  \bibinfo {author} {\bibfnamefont {B.L.}~\bibnamefont {Altshuler}}, \ and\
  \bibinfo {author} {\bibfnamefont {P.}~\bibnamefont {Coleman}},\ }\bibfield
  {title} {\enquote {\bibinfo {title} {Spectroscopic signatures of
  nonequilibrium pairing in atomic {F}ermi gases},}\ }\href@noop {} {\bibfield
  {journal} {\bibinfo  {journal} {Phys. Rev. Lett.}\ }\textbf {\bibinfo
  {volume} {99}},\ \bibinfo {pages} {160402} (\bibinfo {year}
  {2007})}\BibitemShut {NoStop}%
\bibitem [{\citenamefont {Scott}\ \emph {et~al.}(2012)\citenamefont {Scott},
  \citenamefont {Dalfovo}, \citenamefont {Pitaevskii},\ and\ \citenamefont
  {Stringari}}]{Scott2012Rapid}%
  \BibitemOpen
  \bibfield  {author} {\bibinfo {author} {\bibfnamefont {R.G.}~\bibnamefont
  {Scott}}, \bibinfo {author} {\bibfnamefont {F.}~\bibnamefont {Dalfovo}},
  \bibinfo {author} {\bibfnamefont {L.P.}~\bibnamefont {Pitaevskii}}, \ and\
  \bibinfo {author} {\bibfnamefont {S.}~\bibnamefont {Stringari}},\ }\bibfield
  {title} {\enquote {\bibinfo {title} {Rapid ramps across the {BEC-BCS}
  crossover: A route to measuring the superfluid gap},}\ }\href@noop {}
  {\bibfield  {journal} {\bibinfo  {journal} {Phys. Rev. A}\ }\textbf {\bibinfo
  {volume} {86}},\ \bibinfo {pages} {053604} (\bibinfo {year}
  {2012})}\BibitemShut {NoStop}%
\bibitem [{\citenamefont {Bruun}(2014)}]{Bruun2014Long}%
  \BibitemOpen
  \bibfield  {author} {\bibinfo {author} {\bibfnamefont {G.}~\bibnamefont
  {Bruun}},\ }\bibfield  {title} {\enquote {\bibinfo {title} {Long-lived
  {H}iggs mode in a two-dimensional confined {F}ermi system},}\ }\href@noop {}
  {\bibfield  {journal} {\bibinfo  {journal} {Phys. Rev. A}\ }\textbf {\bibinfo
  {volume} {90}},\ \bibinfo {pages} {023621} (\bibinfo {year}
  {2014})}\BibitemShut {NoStop}%
\bibitem [{\citenamefont {Hannibal}\ \emph {et~al.}(2015)\citenamefont
  {Hannibal}, \citenamefont {Kettmann}, \citenamefont {Croitoru}, \citenamefont
  {Vagov}, \citenamefont {Axt},\ and\ \citenamefont
  {Kuhn}}]{hannibal2015quench}%
  \BibitemOpen
  \bibfield  {author} {\bibinfo {author} {\bibfnamefont {S.}~\bibnamefont
  {Hannibal}}, \bibinfo {author} {\bibfnamefont {P.}~\bibnamefont {Kettmann}},
  \bibinfo {author} {\bibfnamefont {M.~D.}\ \bibnamefont {Croitoru}}, \bibinfo
  {author} {\bibfnamefont {A.}~\bibnamefont {Vagov}}, \bibinfo {author}
  {\bibfnamefont {V.~M.}\ \bibnamefont {Axt}}, \ and\ \bibinfo {author}
  {\bibfnamefont {T.}~\bibnamefont {Kuhn}},\ }\bibfield  {title} {\enquote
  {\bibinfo {title} {Quench dynamics of an ultracold {F}ermi gas in the {BCS}
  regime: Spectral properties and confinement-induced breakdown of the {H}iggs
  mode},}\ }\href@noop {} {\bibfield  {journal} {\bibinfo  {journal} {Phys.
  Rev. A}\ }\textbf {\bibinfo {volume} {91}},\ \bibinfo {pages} {043630}
  (\bibinfo {year} {2015})}\BibitemShut {NoStop}%
\bibitem [{\citenamefont {Kinast}\ \emph
  {et~al.}(2004{\natexlab{a}})\citenamefont {Kinast}, \citenamefont {Hemmer},
  \citenamefont {Gehm}, \citenamefont {Turlapov},\ and\ \citenamefont
  {Thomas}}]{Kinast2004Evidence}%
  \BibitemOpen
  \bibfield  {author} {\bibinfo {author} {\bibfnamefont {J.}~\bibnamefont
  {Kinast}}, \bibinfo {author} {\bibfnamefont {S.L.}~\bibnamefont {Hemmer}},
  \bibinfo {author} {\bibfnamefont {M.E.}~\bibnamefont {Gehm}}, \bibinfo {author}
  {\bibfnamefont {A.}~\bibnamefont {Turlapov}}, \ and\ \bibinfo {author}
  {\bibfnamefont {J.E.}~\bibnamefont {Thomas}},\ }\bibfield  {title} {\enquote
  {\bibinfo {title} {Evidence for superfluidity in a resonantly interacting
  {F}ermi gas},}\ }\href@noop {} {\bibfield  {journal} {\bibinfo  {journal}
  {Phys. Rev. Lett.}\ }\textbf {\bibinfo {volume} {92}},\ \bibinfo {pages}
  {150402} (\bibinfo {year} {2004}{\natexlab{a}})}\BibitemShut {NoStop}%
\bibitem [{\citenamefont {Kinast}\ \emph
  {et~al.}(2004{\natexlab{b}})\citenamefont {Kinast}, \citenamefont
  {Turlapov},\ and\ \citenamefont {Thomas}}]{Kinast2004Breakdown}%
  \BibitemOpen
  \bibfield  {author} {\bibinfo {author} {\bibfnamefont {J.}~\bibnamefont
  {Kinast}}, \bibinfo {author} {\bibfnamefont {A.}~\bibnamefont {Turlapov}}, \
  and\ \bibinfo {author} {\bibfnamefont {J.E.}~\bibnamefont {Thomas}},\
  }\bibfield  {title} {\enquote {\bibinfo {title} {Breakdown of hydrodynamics
  in the radial breathing mode of a strongly interacting {F}ermi gas},}\
  }\href@noop {} {\bibfield  {journal} {\bibinfo  {journal} {Phys. Rev. A}\
  }\textbf {\bibinfo {volume} {70}},\ \bibinfo {pages} {051401} (\bibinfo
  {year} {2004}{\natexlab{b}})}\BibitemShut {NoStop}%
\bibitem [{\citenamefont {Bartenstein}\ \emph {et~al.}(2004)\citenamefont
  {Bartenstein}, \citenamefont {Altmeyer}, \citenamefont {Riedl}, \citenamefont
  {Jochim}, \citenamefont {Chin}, \citenamefont {Denschlag},\ and\
  \citenamefont {Grimm}}]{bartenstein2004Collective}%
  \BibitemOpen
  \bibfield  {author} {\bibinfo {author} {\bibfnamefont {M.}~\bibnamefont
  {Bartenstein}}, \bibinfo {author} {\bibfnamefont {A.}~\bibnamefont
  {Altmeyer}}, \bibinfo {author} {\bibfnamefont {S.}~\bibnamefont {Riedl}},
  \bibinfo {author} {\bibfnamefont {S.}~\bibnamefont {Jochim}}, \bibinfo
  {author} {\bibfnamefont {C.}~\bibnamefont {Chin}}, \bibinfo {author}
  {\bibfnamefont {J.~H.}\ \bibnamefont {Denschlag}}, \ and\ \bibinfo {author}
  {\bibfnamefont {R.}~\bibnamefont {Grimm}},\ }\bibfield  {title} {\enquote
  {\bibinfo {title} {Collective excitations of a degenerate gas at the
  {BEC-BCS} crossover},}\ }\href@noop {} {\bibfield  {journal} {\bibinfo
  {journal} {Phys. Rev. Lett.}\ }\textbf {\bibinfo {volume} {92}},\ \bibinfo
  {pages} {203201} (\bibinfo {year} {2004})}\BibitemShut {NoStop}%
\bibitem [{\citenamefont {Altmeyer}\ \emph
  {et~al.}(2007{\natexlab{a}})\citenamefont {Altmeyer}, \citenamefont {Riedl},
  \citenamefont {Wright}, \citenamefont {Kohstall}, \citenamefont {Denschlag},\
  and\ \citenamefont {Grimm}}]{altmeyer2007dynamics}%
  \BibitemOpen
  \bibfield  {author} {\bibinfo {author} {\bibfnamefont {A.}~\bibnamefont
  {Altmeyer}}, \bibinfo {author} {\bibfnamefont {S.}~\bibnamefont {Riedl}},
  \bibinfo {author} {\bibfnamefont {M.~J.}\ \bibnamefont {Wright}}, \bibinfo
  {author} {\bibfnamefont {C.}~\bibnamefont {Kohstall}}, \bibinfo {author}
  {\bibfnamefont {J.~H.}\ \bibnamefont {Denschlag}}, \ and\ \bibinfo {author}
  {\bibfnamefont {R.}~\bibnamefont {Grimm}},\ }\bibfield  {title} {\enquote
  {\bibinfo {title} {Dynamics of a strongly interacting {F}ermi gas: The radial
  quadrupole mode},}\ }\href@noop {} {\bibfield  {journal} {\bibinfo  {journal}
  {Phys. Rev. A}\ }\textbf {\bibinfo {volume} {76}},\ \bibinfo {pages} {033610}
  (\bibinfo {year} {2007}{\natexlab{a}})}\BibitemShut {NoStop}%
\bibitem [{\citenamefont {Altmeyer}\ \emph
  {et~al.}(2007{\natexlab{b}})\citenamefont {Altmeyer}, \citenamefont {Riedl},
  \citenamefont {Kohstall}, \citenamefont {Wright}, \citenamefont {Geursen},
  \citenamefont {Bartenstein}, \citenamefont {Chin}, \citenamefont
  {Denschlag},\ and\ \citenamefont {Grimm}}]{altmeyer2007precision}%
  \BibitemOpen
  \bibfield  {author} {\bibinfo {author} {\bibfnamefont {A.}~\bibnamefont
  {Altmeyer}}, \bibinfo {author} {\bibfnamefont {S.}~\bibnamefont {Riedl}},
  \bibinfo {author} {\bibfnamefont {C.}~\bibnamefont {Kohstall}}, \bibinfo
  {author} {\bibfnamefont {M.~J.}\ \bibnamefont {Wright}}, \bibinfo {author}
  {\bibfnamefont {R.}~\bibnamefont {Geursen}}, \bibinfo {author} {\bibfnamefont
  {M.}~\bibnamefont {Bartenstein}}, \bibinfo {author} {\bibfnamefont
  {C.}~\bibnamefont {Chin}}, \bibinfo {author} {\bibfnamefont {J.~H.}\
  \bibnamefont {Denschlag}}, \ and\ \bibinfo {author} {\bibfnamefont
  {R.}~\bibnamefont {Grimm}},\ }\bibfield  {title} {\enquote {\bibinfo {title}
  {Precision measurements of collective oscillations in the {BEC-BCS}
  crossover},}\ }\href@noop {} {\bibfield  {journal} {\bibinfo  {journal}
  {Phys. Rev. Lett.}\ }\textbf {\bibinfo {volume} {98}},\ \bibinfo {pages}
  {040401} (\bibinfo {year} {2007}{\natexlab{b}})}\BibitemShut {NoStop}%
\bibitem [{\citenamefont {Riedl}\ \emph {et~al.}(2008)\citenamefont {Riedl},
  \citenamefont {Guajardo}, \citenamefont {Kohstall}, \citenamefont {Altmeyer},
  \citenamefont {Wright}, \citenamefont {Denschlag}, \citenamefont {Grimm},
  \citenamefont {Bruun},\ and\ \citenamefont {Smith}}]{riedl2008collective}%
  \BibitemOpen
  \bibfield  {author} {\bibinfo {author} {\bibfnamefont {S.}~\bibnamefont
  {Riedl}}, \bibinfo {author} {\bibfnamefont {E.~R.~S{\'a}nchez}\ \bibnamefont
  {Guajardo}}, \bibinfo {author} {\bibfnamefont {C.}~\bibnamefont {Kohstall}},
  \bibinfo {author} {\bibfnamefont {A.}~\bibnamefont {Altmeyer}}, \bibinfo
  {author} {\bibfnamefont {M.~J.}\ \bibnamefont {Wright}}, \bibinfo {author}
  {\bibfnamefont {J.~H.}\ \bibnamefont {Denschlag}}, \bibinfo {author}
  {\bibfnamefont {R.}~\bibnamefont {Grimm}}, \bibinfo {author} {\bibfnamefont
  {G.~M.}\ \bibnamefont {Bruun}}, \ and\ \bibinfo {author} {\bibfnamefont
  {H.}~\bibnamefont {Smith}},\ }\bibfield  {title} {\enquote {\bibinfo {title}
  {Collective oscillations of a {F}ermi gas in the unitarity limit: Temperature
  effects and the role of pair correlations},}\ }\href@noop {} {\bibfield
  {journal} {\bibinfo  {journal} {Phys. Rev. A}\ }\textbf {\bibinfo {volume}
  {78}},\ \bibinfo {pages} {053609} (\bibinfo {year} {2008})}\BibitemShut
  {NoStop}%
\bibitem [{\citenamefont {Baranov}\ and\ \citenamefont
  {Petrov}(2000)}]{baranov2000low}%
  \BibitemOpen
  \bibfield  {author} {\bibinfo {author} {\bibfnamefont {M.~A.}\ \bibnamefont
  {Baranov}}\ and\ \bibinfo {author} {\bibfnamefont {D.~S.}\ \bibnamefont
  {Petrov}},\ }\bibfield  {title} {\enquote {\bibinfo {title} {Low-energy
  collective excitations in a superfluid trapped {F}ermi gas},}\ }\href@noop {}
  {\bibfield  {journal} {\bibinfo  {journal} {Phys. Rev. A}\ }\textbf {\bibinfo
  {volume} {62}},\ \bibinfo {pages} {041601} (\bibinfo {year}
  {2000})}\BibitemShut {NoStop}%
\bibitem [{\citenamefont {Bruun}\ and\ \citenamefont
  {Mottelson}(2001)}]{bruun2001low}%
  \BibitemOpen
  \bibfield  {author} {\bibinfo {author} {\bibfnamefont {G.~M.}\ \bibnamefont
  {Bruun}}\ and\ \bibinfo {author} {\bibfnamefont {B.~R.}\ \bibnamefont
  {Mottelson}},\ }\bibfield  {title} {\enquote {\bibinfo {title} {Low energy
  collective modes of a superfluid trapped atomic {F}ermi gas},}\ }\href@noop
  {} {\bibfield  {journal} {\bibinfo  {journal} {Phys. Rev. Lett.}\ }\textbf
  {\bibinfo {volume} {87}},\ \bibinfo {pages} {270403} (\bibinfo {year}
  {2001})}\BibitemShut {NoStop}%
\bibitem [{\citenamefont {Bruun}(2002)}]{bruun2002low}%
  \BibitemOpen
  \bibfield  {author} {\bibinfo {author} {\bibfnamefont {G.~M.}\ \bibnamefont
  {Bruun}},\ }\bibfield  {title} {\enquote {\bibinfo {title} {Low-energy
  monopole modes of a trapped atomic {F}ermi gas},}\ }\href@noop {} {\bibfield
  {journal} {\bibinfo  {journal} {Phys. Rev. Lett.}\ }\textbf {\bibinfo
  {volume} {89}},\ \bibinfo {pages} {263002} (\bibinfo {year}
  {2002})}\BibitemShut {NoStop}%
\bibitem [{\citenamefont {Hu}\ \emph {et~al.}(2004)\citenamefont {Hu},
  \citenamefont {Minguzzi}, \citenamefont {Liu},\ and\ \citenamefont
  {Tosi}}]{hu2004collective}%
  \BibitemOpen
  \bibfield  {author} {\bibinfo {author} {\bibfnamefont {H.}~\bibnamefont
  {Hu}}, \bibinfo {author} {\bibfnamefont {A.}~\bibnamefont {Minguzzi}},
  \bibinfo {author} {\bibfnamefont {X.~J.}\ \bibnamefont {Liu}}, \ and\
  \bibinfo {author} {\bibfnamefont {M.~P.}\ \bibnamefont {Tosi}},\ }\bibfield
  {title} {\enquote {\bibinfo {title} {Collective modes and ballistic expansion
  of a {F}ermi gas in the {BCS-BEC} crossover},}\ }\href@noop {} {\bibfield
  {journal} {\bibinfo  {journal} {Phys. Rev. Lett.}\ }\textbf {\bibinfo
  {volume} {93}},\ \bibinfo {pages} {190403} (\bibinfo {year}
  {2004})}\BibitemShut {NoStop}%
\bibitem [{\citenamefont {Heiselberg}(2004)}]{heiselberg2004collective}%
  \BibitemOpen
  \bibfield  {author} {\bibinfo {author} {\bibfnamefont {H.}~\bibnamefont
  {Heiselberg}},\ }\bibfield  {title} {\enquote {\bibinfo {title} {Collective
  modes of trapped gases at the {BEC-BCS} crossover},}\ }\href@noop {}
  {\bibfield  {journal} {\bibinfo  {journal} {Phys. Rev. Lett.}\ }\textbf
  {\bibinfo {volume} {93}},\ \bibinfo {pages} {040402} (\bibinfo {year}
  {2004})}\BibitemShut {NoStop}%
\bibitem [{\citenamefont {Stringari}(2004)}]{Stringari2004Collective}%
  \BibitemOpen
  \bibfield  {author} {\bibinfo {author} {\bibfnamefont {S.}~\bibnamefont
  {Stringari}},\ }\bibfield  {title} {\enquote {\bibinfo {title} {Collective
  oscillations of a trapped superfluid {F}ermi gas near a {F}eshbach
  resonance},}\ }\href@noop {} {\bibfield  {journal} {\bibinfo  {journal}
  {Europhys. Lett.}\ }\textbf {\bibinfo {volume} {65}},\ \bibinfo {pages}
  {749--752} (\bibinfo {year} {2004})}\BibitemShut {NoStop}%
\bibitem [{\citenamefont {Grasso}\ \emph {et~al.}(2005)\citenamefont {Grasso},
  \citenamefont {Khan},\ and\ \citenamefont {Urban}}]{grasso2005temperature}%
  \BibitemOpen
  \bibfield  {author} {\bibinfo {author} {\bibfnamefont {M.}~\bibnamefont
  {Grasso}}, \bibinfo {author} {\bibfnamefont {E.}~\bibnamefont {Khan}}, \ and\
  \bibinfo {author} {\bibfnamefont {M.}~\bibnamefont {Urban}},\ }\bibfield
  {title} {\enquote {\bibinfo {title} {Temperature dependence and finite-size
  effects in collective modes of superfluid-trapped {F}ermi gases},}\
  }\href@noop {} {\bibfield  {journal} {\bibinfo  {journal} {Phys. Rev. A}\
  }\textbf {\bibinfo {volume} {72}},\ \bibinfo {pages} {043617} (\bibinfo
  {year} {2005})}\BibitemShut {NoStop}%
\bibitem [{\citenamefont {Korolyuk}\ \emph {et~al.}(2011)\citenamefont
  {Korolyuk}, \citenamefont {Kinnunen},\ and\ \citenamefont
  {T{\"o}rm{\"a}}}]{korolyuk2011density}%
  \BibitemOpen
  \bibfield  {author} {\bibinfo {author} {\bibfnamefont {A.}~\bibnamefont
  {Korolyuk}}, \bibinfo {author} {\bibfnamefont {J.~J.}\ \bibnamefont
  {Kinnunen}}, \ and\ \bibinfo {author} {\bibfnamefont {P.}~\bibnamefont
  {T{\"o}rm{\"a}}},\ }\bibfield  {title} {\enquote {\bibinfo {title} {Density
  response of a trapped {F}ermi gas: A crossover from the pair vibration mode
  to the {G}oldstone mode},}\ }\href@noop {} {\bibfield  {journal} {\bibinfo
  {journal} {Phys. Rev. A}\ }\textbf {\bibinfo {volume} {84}},\ \bibinfo
  {pages} {033623} (\bibinfo {year} {2011})}\BibitemShut {NoStop}%
\bibitem [{\citenamefont {Clark}\ \emph {et~al.}(2015)\citenamefont {Clark},
  \citenamefont {Ha}, \citenamefont {Xu},\ and\ \citenamefont
  {Chin}}]{clark2015quantum}%
  \BibitemOpen
  \bibfield  {author} {\bibinfo {author} {\bibfnamefont {L.~W.}\ \bibnamefont
  {Clark}}, \bibinfo {author} {\bibfnamefont {L.~C.}\ \bibnamefont {Ha}},
  \bibinfo {author} {\bibfnamefont {C.~Y.}\ \bibnamefont {Xu}}, \ and\ \bibinfo
  {author} {\bibfnamefont {C.}~\bibnamefont {Chin}},\ }\bibfield  {title}
  {\enquote {\bibinfo {title} {Quantum dynamics with spatiotemporal control of
  interactions in a stable {B}ose-{E}instein condensate},}\ }\href@noop {}
  {\bibfield  {journal} {\bibinfo  {journal} {Phys. Rev. Lett.}\ }\textbf
  {\bibinfo {volume} {115}},\ \bibinfo {pages} {155301} (\bibinfo {year}
  {2015})}\BibitemShut {NoStop}%
\bibitem [{\citenamefont {De~Gennes}(1989)}]{DeGennes1989Superconductivity}%
  \BibitemOpen
  \bibfield  {author} {\bibinfo {author} {\bibfnamefont {P.}~\bibnamefont
  {De~Gennes}},\ }\href@noop {} {\emph {\bibinfo {title} {Superconductivity of
  metals and alloys}}}\ (\bibinfo  {publisher} {Addison-Wesley New York},\
  \bibinfo {year} {1989})\BibitemShut {NoStop}%
\bibitem [{\citenamefont {Datta}\ and\ \citenamefont
  {Bagwell}(1999)}]{Datta1999Can}%
  \BibitemOpen
  \bibfield  {author} {\bibinfo {author} {\bibfnamefont {S.}~\bibnamefont
  {Datta}}\ and\ \bibinfo {author} {\bibfnamefont {P.~F.}\ \bibnamefont
  {Bagwell}},\ }\bibfield  {title} {\enquote {\bibinfo {title} {Can the
  {B}ogoliubov-de {G}ennes equation be interpreted as a one-particle wave
  equation?}}\ }\href@noop {} {\bibfield  {journal} {\bibinfo  {journal}
  {Superlattices and Microstruct.}\ }\textbf {\bibinfo {volume} {25}},\
  \bibinfo {pages} {1233--1250} (\bibinfo {year} {1999})}\BibitemShut {NoStop}%
\bibitem [{\citenamefont {Lord}(1949)}]{lord1949some}%
  \BibitemOpen
  \bibfield  {author} {\bibinfo {author} {\bibfnamefont {R.~D.}\ \bibnamefont
  {Lord}},\ }\bibfield  {title} {\enquote {\bibinfo {title} {Some integrals
  involving {H}ermite polynomials},}\ }\href@noop {} {\bibfield  {journal}
  {\bibinfo  {journal} {Journal of the London Mathematical Society}\ }\textbf
  {\bibinfo {volume} {1}},\ \bibinfo {pages} {101--112} (\bibinfo {year}
  {1949})}\BibitemShut {NoStop}%
\bibitem [{\citenamefont {Shanenko}\ \emph {et~al.}(2012)\citenamefont
  {Shanenko}, \citenamefont {Croitoru}, \citenamefont {Vagov}, \citenamefont
  {Axt}, \citenamefont {Perali},\ and\ \citenamefont
  {Peeters}}]{Shanenko2012Atypical}%
  \BibitemOpen
  \bibfield  {author} {\bibinfo {author} {\bibfnamefont {A.A.}~\bibnamefont
  {Shanenko}}, \bibinfo {author} {\bibfnamefont {M.~D.}\ \bibnamefont
  {Croitoru}}, \bibinfo {author} {\bibfnamefont {A.V.}~\bibnamefont {Vagov}},
  \bibinfo {author} {\bibfnamefont {V.M.}~\bibnamefont {Axt}}, \bibinfo {author}
  {\bibfnamefont {A.}~\bibnamefont {Perali}}, \ and\ \bibinfo {author}
  {\bibfnamefont {F.M.}~\bibnamefont {Peeters}},\ }\bibfield  {title} {\enquote
  {\bibinfo {title} {Atypical {BCS-BEC} crossover induced by quantum-size
  effects},}\ }\href@noop {} {\bibfield  {journal} {\bibinfo  {journal} {Phys.
  Rev. A}\ }\textbf {\bibinfo {volume} {86}},\ \bibinfo {pages} {033612}
  (\bibinfo {year} {2012})}\BibitemShut {NoStop}%
\bibitem [{\citenamefont {Jain}(2017)}]{Jain2017Theory}%
  \BibitemOpen
  \bibfield  {author} {\bibinfo {author} {\bibfnamefont {Akash}\ \bibnamefont
  {Jain}},\ }\bibfield  {title} {\enquote {\bibinfo {title} {Theory of
  non-abelian superfluid dynamics},}\ }\href {\doibase
  10.1103/PhysRevD.95.121701} {\bibfield  {journal} {\bibinfo  {journal} {Phys.
  Rev. D}\ }\textbf {\bibinfo {volume} {95}},\ \bibinfo {pages} {121701}
  (\bibinfo {year} {2017})}\BibitemShut {NoStop}%
\bibitem [{\citenamefont {Yuzbashyan}\ \emph {et~al.}(2015)\citenamefont
  {Yuzbashyan}, \citenamefont {Dzero}, \citenamefont {Gurarie},\ and\
  \citenamefont {Foster}}]{Yuzbashyan2015Quantum}%
  \BibitemOpen
  \bibfield  {author} {\bibinfo {author} {\bibfnamefont {Emil~A.}\ \bibnamefont
  {Yuzbashyan}}, \bibinfo {author} {\bibfnamefont {Maxim}\ \bibnamefont
  {Dzero}}, \bibinfo {author} {\bibfnamefont {Victor}\ \bibnamefont {Gurarie}},
  \ and\ \bibinfo {author} {\bibfnamefont {Matthew~S.}\ \bibnamefont
  {Foster}},\ }\bibfield  {title} {\enquote {\bibinfo {title} {Quantum quench
  phase diagrams of an s-wave {BCS-BEC} condensate},}\ }\href@noop {}
  {\bibfield  {journal} {\bibinfo  {journal} {Physical Review A}\ }\textbf
  {\bibinfo {volume} {91}},\ \bibinfo {pages} {033628} (\bibinfo {year}
  {2015})}\BibitemShut {NoStop}%
\bibitem [{\citenamefont {Stewart}\ \emph {et~al.}(2008)\citenamefont
  {Stewart}, \citenamefont {Gaebler},\ and\ \citenamefont
  {Jin}}]{Stewart2008Using}%
  \BibitemOpen
  \bibfield  {author} {\bibinfo {author} {\bibfnamefont {J.}~\bibnamefont
  {Stewart}}, \bibinfo {author} {\bibfnamefont {J.}~\bibnamefont {Gaebler}}, \
  and\ \bibinfo {author} {\bibfnamefont {D.}~\bibnamefont {Jin}},\ }\bibfield
  {title} {\enquote {\bibinfo {title} {Using photoemission spectroscopy to
  probe a strongly interacting {F}ermi gas},}\ }\href@noop {} {\bibfield
  {journal} {\bibinfo  {journal} {Nature}\ }\textbf {\bibinfo {volume} {454}},\
  \bibinfo {pages} {744--747} (\bibinfo {year} {2008})}\BibitemShut {NoStop}%
\bibitem [{\citenamefont {Stewart}\ \emph {et~al.}(2010)\citenamefont
  {Stewart}, \citenamefont {Gaebler}, \citenamefont {Drake},\ and\
  \citenamefont {Jin}}]{Stewart2010Verification}%
  \BibitemOpen
  \bibfield  {author} {\bibinfo {author} {\bibfnamefont {J.T.}~\bibnamefont
  {Stewart}}, \bibinfo {author} {\bibfnamefont {J.P.}~\bibnamefont {Gaebler}},
  \bibinfo {author} {\bibfnamefont {T.E.}~\bibnamefont {Drake}}, \ and\ \bibinfo
  {author} {\bibfnamefont {D.S.}~\bibnamefont {Jin}},\ }\bibfield  {title}
  {\enquote {\bibinfo {title} {Verification of universal relations in a
  strongly interacting {F}ermi gas},}\ }\href@noop {} {\bibfield  {journal}
  {\bibinfo  {journal} {Phys. Rev. Lett.}\ }\textbf {\bibinfo {volume} {104}},\
  \bibinfo {pages} {235301} (\bibinfo {year} {2010})}\BibitemShut {NoStop}%
\bibitem [{\citenamefont {Ketterle}\ and\ \citenamefont
  {Zwierlein}(2008)}]{Ketterle2008Making}%
  \BibitemOpen
  \bibfield  {author} {\bibinfo {author} {\bibfnamefont {W.}~\bibnamefont
  {Ketterle}}\ and\ \bibinfo {author} {\bibfnamefont {M.}~\bibnamefont
  {Zwierlein}},\ }\bibfield  {title} {\enquote {\bibinfo {title} {Making,
  probing and understanding ultracold {F}ermi gases},}\ }\href@noop {}
  {\bibfield  {journal} {\bibinfo  {journal} {ArXiv}\ } (\bibinfo {year}
  {2008})}\BibitemShut {NoStop}%
\end{thebibliography}

%

\end{document}